\documentclass[prd,aps,showpacs]{revtex4}
\usepackage{amssymb,graphicx}

\newcommand{\re}{\mbox{Re}}
\newcommand{\im}{\mbox{Im}}

\newcommand{\Ez}{\mbox{\em \r{E}\hspace{0.3mm}}}
\newcommand{\Hz}{\mbox{\em \r{H}\hspace{0.3mm}}}

\newcommand{\Natural}{\mathbb N}
\newcommand{\Real}{\mathbb{R}}

\newcommand{\identy}{1\!\!1}
\newcommand{\hateq}{\; \hat{=}\; }

\begin{document}

\title{Towards absorbing outer boundaries in General Relativity}

\date{\today}
\author{Luisa T. Buchman$^{1,2}$ and Olivier C. A. Sarbach$^{2,3,4}$}
\affiliation{$^1$Relativistic Astrophysics 169-327,
Jet Propulsion Laboratory,
4800 Oak Grove Drive,
Pasadena, California 91109--8099, USA}
\affiliation{$^2$Theoretical Astrophysics 130-33,
California Institute of Technology,
1200 East California Boulevard,
Pasadena, California 91125--0001, USA}
\affiliation{$^3$Department of Mathematics, 
University of California at San Diego, 9500 Gilman Drive, La Jolla, 
California 92093--0112, USA}
\affiliation{$^4$Instituto de F\a'{\i}sica y Matem\a'aticas,
Universidad Michoacana de San Nicol\a'as de Hidalgo, Edificio C-3, 
Cd. Universitaria. C. P. 58040 Morelia, Michoac\a'an, M\a'exico}

\begin{abstract}
We construct exact solutions to the Bianchi equations on a flat
spacetime background. When the constraints are satisfied, these
solutions represent in- and outgoing linearized gravitational
radiation. We then consider the Bianchi equations on a subset of flat
spacetime of the form $[0,T] \times B_R$, where $B_R$ is a ball of
radius $R$, and analyze different kinds of boundary conditions on
$\partial B_R$. Our main results are: i) We give an explicit analytic
example showing that boundary conditions obtained from freezing the
incoming characteristic fields to their initial values are not
compatible with the constraints. ii) With the help of the exact
solutions constructed, we determine the amount of artificial
reflection of gravitational radiation from constraint-preserving
boundary conditions which freeze the Weyl scalar $\Psi_0$ to its
initial value. For monochromatic radiation with wave number $k$ and
arbitrary angular momentum number $\ell\geq 2$, the amount of
reflection decays as $(k R)^{-4}$ for large $k R$. iii) For each $L
\geq 2$, we construct new local constraint-preserving boundary
conditions which perfectly absorb linearized radiation with $\ell \leq
L$. (iv) We generalize our analysis to a weakly curved background of
mass $M$, and compute first order corrections in $M/R$ to the
reflection coefficients for quadrupolar odd-parity radiation. For our
new boundary condition with $L=2$, the reflection coefficient is
smaller than the one for the freezing $\Psi_0$ boundary condition by a
factor of $M/R$ for $k R > 1.04$. Implications of these results for
numerical simulations of binary black holes on finite domains are
discussed.
\end{abstract}


\maketitle

\section{Introduction}
\label{Sect:Intro}

A common approach for numerically solving the Einstein field equations
on a spatially unbounded domain is to truncate the domain via an
artificial boundary, thus forming a finite computational domain
$\Omega$ with outer boundary $\partial\Omega$\footnote{If the
spacetime contains black holes with excised singularities, $\Omega$
will also possess inner boundaries.}. In order to obtain a unique
Cauchy evolution, it is necessary to impose boundary conditions at
$\partial \Omega$. These boundary conditions should form a well posed
initial boundary value problem (IBVP) and, ideally, be completely
transparent to the physical problem on the unbounded domain. Short of
achieving the ideal, one can try to develop so-called absorbing
boundary conditions which form a well posed IBVP and insure that only
a very small amount of spurious gravitational radiation is reflected
from $\partial\Omega$ into the computational domain. Once the IBVP on
$\Omega$ is formulated, it is solved via a numerical approximation
scheme which, together with the truncation of the domain, introduces
two artificial parameters: a discretization parameter $h$, describing
the coarseness of the discretization, and a cut-off parameter $R$,
which gives the size of the spatial domain $\Omega$. For a stable
discretization, it is expected that the continuum solution of the
unbounded problem is recovered in the limit where $h \to 0$ and $R \to
\infty$. In practice, due to finite computer resources, it is not
possible to take this limit. Instead, one needs to quantify how small
$h$ and how large $R$ need to be so that the error is below a certain
tolerance value.

In this article, we address the ``$R$-dependent'' part of this
task. We analyze boundary conditions which have been recently
presented in the literature, and provide estimates for the amount of
spurious radiation coming from $\partial\Omega$. Additionally, we
propose new boundary conditions for Einstein's vacuum field equations
which introduce significantly less reflections than existing conditions.

There has been a substantial amount of work on the construction of
absorbing (also called non-reflecting in the literature) boundary
conditions for wave problems in acoustics, electromagnetism,
meteorology, and solid geophysics (see \cite{dG91} for a review). One
approach is based on a sequence of {\em local} boundary conditions
\cite{bEaM77,aBeT80,rH86} with increasing order of accuracy. Although
higher order local boundary conditions usually involve solving a high
order differential equation at the boundary, the problem can be dealt
with by introducing auxiliary variables at the boundary surface
\cite{dG01,dGbN03}. A different approach is based on fast converging
series expansions of {\em exact nonlocal} boundary conditions (see
\cite{bAlGtH2000} and references therein). Of particular interest for
this article is the work by Lau \cite{sL04a,sL04b,sL05}, which
generalizes the work in Ref. \cite{bAlGtH2000} to the construction of
exact non-reflecting boundary conditions for the Regge-Wheeler and
Zerilli equations, describing linear gravitational fluctuations about
a Schwarzschild black hole. This approach is robust, very accurate,
and stable. However, it is based on a detailed knowledge of the
solutions which might not always be available in more general
situations.

For the fully nonlinear Einstein equations, the construction of
absorbing outer boundary conditions is particularly difficult. First
of all, Einstein's field equations determine the evolution of the
metric tensor, so one does not know the geometrical structure of the
spacetime before actually solving the IBVP. Hence, it is not clear
{\em a priori} how the geometry of the outer boundary evolves. This
poses a problem if one wants to fix, for example, the area of the
boundary $\partial\Omega$ to its initial value. Second, in the Cauchy
formulation of Einstein's field equations, there exist
constraint-violating modes which propagate with nontrivial
characteristic speeds. This is in contrast to the standard Cauchy
formulation of Maxwell's equations, where the evolution equations
imply that the constraint variables (namely, the divergence of the
electric and magnetic fields) are constant in time. Since the
constraint variables in General Relativity propagate non-trivially,
constraint-preserving boundary conditions (CPBC) must be specified so
that constraint violations are not introduced into the computational
domain. Finally, in General Relativity, it is difficult to define
precisely what is meant by outgoing and ingoing radiation. This is due
to the nonlinear nature of the theory and its diffeomorphism
invariance. (See Ref. \cite{jBlB02} for a discussion of this problem
for nonlinear gravitational plane waves.) These issues all contribute
to the challenge of determining the amount of spurious reflections
from the outer boundary.

A significant advance towards developing absorbing boundary conditions
for General Relativity was the first (and, to date, the only) well
posed IBVP for Einstein's vacuum field equations presented in
Ref. \cite{hFgN99}. This work, which is based on a tetrad formulation,
recasts the evolution equations into a first order symmetric
hyperbolic form with maximally dissipative boundary conditions, for
which (local in time) well posedness is guaranteed \cite{pS96b}. The
boundary conditions constructed in \cite{hFgN99} control part of the
geometry of the boundary surface by specifying its constant mean
curvature, control the radiation by prescribing suitable combinations
of the complex Newman-Penrose scalars $\Psi_0$ and $\Psi_4$, where the
null tetrad is constructed from the evolution vector field and the
normal to the boundary, and are constraint-preserving. Recently, there
has been considerable effort to generalize the work in
Ref. \cite{hFgN99} by specifying CPBC for the more commonly used
metric formulations of gravity (see
Refs. \cite{bSjW03,gCjPoRoSmT03,oSmT05,lKlLmSlBhP05,lLmSlKrOoR06,
gNoS06,hKjW06,oR06} and references therein). In particular, the
methods in Refs. \cite{lKlLmSlBhP05,oSmT05,gNoS06,lLmSlKrOoR06,oR06}
in addition to preserving the constraints, regulate the dynamical
degrees of freedom by freezing the Newman-Penrose scalar $\Psi_0$,
defined with respect to a suitably chosen null tetrad, to its initial
value\footnote{Actually, the formulations in
Refs. \cite{hFgN99,oSmT05,gNoS06} also consider more general boundary
conditions which allow to couple $\Psi_4$ to $\Psi_0$.}.

Work focused on eliminating reflections from the outer boundary during
fully relativistic vacuum simulations has been performed by several
authors. In Ref. \cite{jNsB04}, boundary conditions based on the work
in Ref. \cite{aBeT80}, which are perfectly absorbing for quadrupolar
solutions of the flat wave equation, are numerically implemented via
spectral methods, and used in a constrained evolution scheme of
Einstein's field equations \cite{sBeGpG04}. In
Refs. \cite{aAcE88,aAcE90}, solutions of the full nonlinear Einstein
equations on a finite computational domain are matched to exact
analytic, purely outgoing solutions of the weak field equations at the
outer boundary of the
domain. Refs. \cite{aAetal98,mRaAlR98,lRaArMmRsS99,bZePpDmT06}
generalize this idea by matching the nonlinear equations to an ``outer
module'', a code in which the equations are linearized about a
Schwarzschild background, in order to carry the waveforms far into the
wave zone. However, at the interface where the matching occurs, the
methods in Refs. \cite{aAcE88,aAcE90,aAetal98,mRaAlR98,lRaArMmRsS99}
do not take into account either the constraints of the nonlinear
Cauchy code or the characteristic structure of the nonlinear evolution
equations, so it is not clear if the resulting problem is well posed
(the work in \cite{bZePpDmT06}, on the other hand, does take into
account the constraints and the characteristic fields of the Cauchy
code, and gives an implementation for the spherically symmetric Einstein
equations coupled to a massless scalar field). Two other approaches
presented in the literature for constructing absorbing boundary
conditions are: matching the nonlinear Cauchy code to a nonlinear
characteristic code (see Ref. \cite{jW01} for a review and \cite{gC06}
for recent work) and matching an incoming characteristic formulation
to an outgoing one at a time-like cylinder \cite{lL00}. Finally,
methods that avoid introducing an artificial outer boundary altogether
compactify spatial infinity \cite{mClLiOrPfPhV03,fP05}, or make use of
hyperboloidal slices and compactify null infinity (see, for instance,
\cite{hF81,jF98,sHcStVaZ05}).

In this article, we take a step closer to the construction of
absorbing boundary conditions in General Relativity. In order to do
so, the IBVP of Einstein's field equations is analyzed on a compact
domain $\Omega \subset\Real^3$ with smooth outer boundary
$\partial\Omega$ and two simplifying assumptions. The first assumption
is that at all times, the boundary surface $\partial\Omega$ is far
from the strong field region, so that the gravitational field near the
outer boundary is weak. As a consequence, the field equations can be
linearized to a first approximation about flat spacetime in the
vicinity of the outer boundary. The linearized field equations can be
conveniently described by the Bianchi equations, which yield a
Lorentz-invariant system for the linearized Weyl tensor having a
structure which is very similar to that of Maxwell's
equations. Moreover, the linearized Weyl tensor is invariant with
respect to infinitesimal coordinate transformations, since it vanishes
on the background \cite{jSmW74}, so there are no gauge modes. The
second assumption is that the boundary $\partial\Omega$ is
approximately a metric sphere of area $4\pi R^2$.  This assumption is
quite natural. In fact, modern numerical relativity codes based on
multi-block finite differencing \cite{jT04,lLoRmT05,eSpDeDmT06} or
pseudo-spectral methods \cite{KST,sBeGpG04} are designed to handle
spherical outer boundaries.

Under these assumptions, it is sufficient to analyze the Bianchi
equations on a domain $\Omega = B_R$ consisting of a ball of radius
$R$. We can then conveniently expand the linearized Weyl tensor in
terms of spherical tensor harmonics, because of the spherical symmetry
of $B_R$. The resulting equations are decoupled: they are a family of
partial differential equations in one spatial dimension parameterized
by the angular momentum numbers $\ell$ and $m$. For each fixed $\ell$
and $m$, the purely dynamical degrees of freedom can be described by a
master equation for the Newman-Penrose scalar $\Psi_2$. This equation
admits exact solutions which propagate along either in- or outgoing
null radial geodesics. The in- and outgoing solutions are related to
each other by a time reversal symmetry $t\mapsto -t$, making it
possible to define sensibly in- and outgoing gravitational
radiation. Hence, in our setting, it is clear how to quantify the
amount of spurious radiation reflected at $\partial B_R$. Using these
exact solutions, we analyze the quality of boundary conditions which
have been proposed in the literature; namely, those which freeze all
the incoming characteristic fields to their initial values, and CPBC
which freeze the Weyl scalar $\Psi_0$ to its initial value.
Furthermore, we offer a set of improved CPBC, which are perfectly
absorbing for linearized radiation on a Minkowski background up to
some arbitrary multipole number $\ell$. Finally, we extend our
analysis to a weakly curved background.

Our main results are the following. First, we show that the naive
boundary condition which freezes all the incoming characteristic
fields to their initial values is not compatible with the
constraints. To show this, we construct explicit solutions to the IBVP
which have the property that they satisfy the constraints exactly on
the initial time slice $t=0$, but violate them at later times $t >
0$. Second, we impose CPBC and freeze the Weyl scalar $\Psi_0$ at the
boundary to its initial value. The exact outgoing solutions do not
satisfy this boundary condition exactly. Specifically, the quantity
$\Psi_0$ constructed from these solutions falls off as $1/r^5$ along
the null geodesics $t=r+{\text const.}$, where $r$ denotes the areal
radius coordinate. This means that a solution to the IBVP
corresponding to the boundary condition $\partial_t\Psi_0 = 0$
consists of a superposition of an in- and an outgoing wave, where the
magnitude of the ratio of the ingoing to the outgoing wave amplitudes
(which we define as the reflection coefficient) measures the amount of
spurious reflection. We find that for monochromatic radiation with
wave number $k$ and arbitrary angular momentum number $\ell\geq 2$,
the reflection coefficient decays as $(kR)^{-4}$ for large $k R$. In
particular, the reflection coefficient lies below $0.1\%$ for
quadrupolar radiation with $k R \geq 6.4$. Third, for each $L \geq 1$,
we construct {\em local} CPBC ${\cal B}_L$ which, for $L \geq 2$,
improve the CPBC involving $\partial_t\Psi_0 = 0$, being {\em
perfectly absorbing} for linearized gravitational radiation on
Minkowski space with angular momentum number $\ell\leq L$. (${\cal
B}_1$ is just the freezing $\Psi_0$ boundary condition and there is no
improvement.)  Since in many practical situations one expects the few
lower multipoles to dominate, an implementation of ${\cal B}_L$ for
$L=2$, $3$, or $4$ should result in only a small amount of spurious
reflection. For $L=2$, our improved boundary condition ${\cal B}_2$
reads
\begin{equation}
\partial_t\left. (\partial_t + \partial_r)( r^5\Psi_0 ) \right|_{r=R} = 0.
\label{Eq:ImprovedBCL=2}
\end{equation}
Finally, we take into account first order corrections from the
curvature of the background. Since we assume that the outer boundary
lies in the weak field regime, we describe spacetime near the outer
boundary by a perturbed Schwarzschild metric of mass $M$, thereby
generalizing our previous analysis by taking into account curvature
near the outer boundary. To estimate the effects due to curvature, we
compute the first order corrections in $2M/R$ to the exact in- and
outgoing solutions with $\ell=2$ and odd parity, and then recalculate
our reflection coefficient for the CPBC involving $\partial_t\Psi_0 =
0$. We find that for $2M/R \ll 1$, the corrected $\ell=2$ odd-parity
reflection coefficient depends only weakly on $2M/R$. In fact, our
results indicate that the reflection coefficient even {\em decreases}
when $2M/R$ increases (but stays small). For quadrupolar solutions
satisfying the improved boundary condition (\ref{Eq:ImprovedBCL=2}),
which is perfectly absorbing for $M=0$, we find that the reflection
coefficient decays as $(2M/R)(kR)^{-4}$ for large $kR$ and small
$2M/R$. More precisely, the reflection coefficient is less than the
one for the freezing $\Psi_0$ boundary condition by a factor of $M/R$
for $k R > 1.04$.

This work is organized as follows. In
Sect. \ref{Sect:BianchiIdentities}, we write down the Bianchi
equations on an arbitrary spacetime. These equations can be obtained
from the Bianchi identities after imposing the Einstein field
equations. Next, we assume the existence of a spacelike foliation and
a preferred radial direction in each timeslice and perform a $2+1+1$
split of the Bianchi equations, which separate into evolution and
constraint equations. The constraint propagation system describing the
evolution of constraint errors is also discussed.

In Sect. \ref{Sect:Exact}, we specialize to a flat spacetime
background of the form $[0,T] \times B_R$, where $B_R$ denotes a ball
of radius $R$. By performing a decomposition into spherical tensor
harmonics with angular momentum number $\ell$, we show that for each
$\ell\geq 2$, the Bianchi equations can be reduced to two master
equations. The first master equation describes the propagation of
constraint violations, and is homogeneous. The second is an equation
for $\Psi_2$, describing the propagation of linearized gravitational
radiation, and has a source term which depends on the solution of the
first master equation. One of the advantages of working with a master
equation for $\Psi_2$ instead of a master equation for $\Psi_0$ or
$\Psi_4$, as is usually done when studying perturbations of black
holes with a Petrov type D metric \cite{jBwP73,Teukolsky72}, is that
for linearization about Minkowski spacetime, the former is invariant
with respect to time reversal. Consequently, there is a nice symmetry
between in- and outgoing solutions: one can be obtained from the other
by changing the sign of $t$. This symmetry makes it possible to define
the reflection coefficients in a natural way. In contrast, under time
reversal, $\Psi_0$ is mapped to conjugate $\Psi_4$ and vice versa, so
that the in- and outgoing parts of $\Psi_0$ look quite different. It
is shown in this section that the master equations governing the
constraint violations and the gravitational radiation both admit exact
analytical solutions, which can be obtained by applying suitable
differential operators to the solution of the one-dimensional flat
wave equation. These solutions can be split in a unique way into in-
and outgoing solutions describing, respectively, in- and outgoing
constraint violations or in- and outgoing gravitational radiation.

In Sect. \ref{Sect:IBVPSol}, we use the exact in- and outgoing
solutions found in the previous section to construct exact solutions
to the IBVP on $B_R$ corresponding to different boundary conditions on
$\partial B_R$. In Sect. \ref{SubSect:Freezing}, we start by analyzing
the characteristic structure of the evolution equations, and specify
boundary conditions which freeze the incoming fields to their initial
values. The incoming fields are related to the Weyl scalars $\Psi_0$
and $\Psi_1$, so these boundary conditions freeze $\Psi_0$ and
$\Psi_1$ at the outer boundary to their initial values. By
constructing an explicit solution with constraint satisfying data at
$t=0$, we show that these ``freezing'' boundary conditions are not
compatible with the constraints in the sense that the solution
violates the constraints for $t > 0$. Next, in
Sect. \ref{SubSect:CPBCFreezingPsi0}, we replace the boundary
condition which freezes $\Psi_1$ to its initial value with CPBC which
guarantee that solutions of the IBVP satisfy the constraints
everywhere on $B_R$ and at all times $t > 0$, provided they hold
initially. This can be achieved in two ways. The first is the one
proposed in Ref. \cite{hFgN99}, which adds suitable combinations of
the constraint equations to the evolution equations so that at the
boundary, the constraints propagate tangentially to the boundary. The
second is to analyze the characteristic structure of the constraint
propagation system, and set the incoming constraint fields to zero at
the boundary. Assuming in what follows that the constraints are
satisfied exactly, we consider only the homogeneous master equation
for $\Psi_2$. We impose the freezing boundary condition
$\partial_t\Psi_0 = 0$ at the boundary $r=R$ on a superposition of in-
and outgoing monochromatic waves for arbitrary $\ell$, and calculate
the resulting reflection coefficients. These coefficients, which
depend only on the dimensionless quantity $k R$ (where $k$ is the wave
number) are of order unity if $k R < \ell$, and decay as $(k R)^{-4}$
for large $k R$. In Sect. \ref{SubSect:CPBCImproved}, we construct the
hierarchy ${\cal B}_1$, ${\cal B}_2$,... of improved boundary
conditions, having the property that ${\cal B}_L$ is perfectly
absorbing for all linearized gravitational waves with angular momentum
number $\ell$ up to and including $L$. The construction of these
boundary conditions is strongly related to the hierarchy proposed in
\cite{aBeT80}.

Finally, in Sect. \ref{Sect:Backscattering}, we generalize our
analysis to odd-parity perturbations of a Schwarzschild background of
mass $M$. Assuming that $M/R \ll 1$, we compute first order
corrections in $M/R$ to the reflection coefficient corresponding to
the freezing $\Psi_0$ boundary condition, for $\ell=2$. In addition,
we compute the reflection coefficient for the boundary condition
${\cal B}_2$ (which is perfectly absorbing for $M=0$), and show that
it is smaller than the one for the freezing $\Psi_0$ condition by a
factor of $M/R$ for $k R > 1.04$.
 
Implications for the modeling of isolated systems such as a binary
black holes are discussed in the conclusions. In an appendix, we show
that the IBVP corresponding to the master equation for $\Psi_2$ and
our new boundary conditions ${\cal B}_2$, ${\cal B}_3$,... is stable
in the sense that the solutions depend uniquely and continuously on
the initial data.

\section{The Bianchi identities}
\label{Sect:BianchiIdentities}

We consider the Bianchi equations
\begin{equation}
\nabla_a C^{a}{}_{bcd} = J_{bcd}\;
\label{Eq:BianchiEq}
\end{equation}
on a given background geometry $(M,g_{ab})$, with $C_{abcd}$ a
tensor field possessing the same algebraic symmetries as the Weyl
tensor:
\begin{equation}
C_{[abc]d} = 0, \qquad
C_{[ab][cd]} = C_{abcd} = C_{cdab}\, ,\qquad
g^{bd} C_{abcd} = 0,
\label{Eq:WeylSym}
\end{equation}
and $J_{bcd}$ a given source tensor which is traceless and satisfies
$J_{[bcd]} = J_{b[cd]} = J_{bcd}$\footnote{Throughout this article,
the indices $a$, $b$, $c$, $d$, $e$, $f$ are spacetime abstract
indices.}. Eq. (\ref{Eq:BianchiEq}) has its origin in the Bianchi
identities,
\begin{equation}
\nabla_a W^{a}{}_{bcd} 
 = \nabla_{[c}\left( G_{d]b} - \frac{1}{3} g_{d]b} g^{ef} G_{ef}\right),
\label{Eq:BianchiId}
\end{equation}
where $W_{abcd}$ and $G_{ab}$ denote, respectively, the Weyl tensor
and the Einstein tensor belonging to the metric $g_{ab}$. If
Einstein's equations are imposed, then the right-hand side of the
identity (\ref{Eq:BianchiId}) can be re-expressed in terms of the
stress-energy tensor, and (\ref{Eq:BianchiId}) becomes an equation for
the Weyl tensor which is of the form of Eq. (\ref{Eq:BianchiEq}). The
identity
\begin{displaymath}
\nabla_a \nabla_b C^{ab}{}_{cd} = C^{ab}{}_{e[c} R^{e}{}_{d]ab}\; ,
\end{displaymath}
where $R_{abcd}$ denotes the Riemann tensor belonging to the
background metric $g_{ab}$, yields the integrability condition
\begin{equation}
\nabla^b J_{bcd} = C^{ab}{}_{e[c} R^{e}{}_{d]ab} \; .
\label{Eq:IntCond}
\end{equation}
The right-hand side of this equation vanishes if $g_{ab}$ is flat or
conformally flat.

In Sections \ref{Sect:Exact} and \ref{Sect:IBVPSol} we will assume
that the background geometry $(M,g_{ab})$ is flat, in which case
Eq. (\ref{Eq:BianchiEq}) describes the propagation of linearized
gravitational radiation, with $C_{abcd}$ the linearized Weyl
tensor. Since the Weyl tensor vanishes for flat spacetime, $C_{abcd}$
is invariant with respect to infinitesimal coordinate transformations
\cite{jSmW74}. As a consequence, the Bianchi equations are well-suited
for studying linearized gravitational waves since they are manifestly
gauge-invariant. For a flat background geometry, the integrability
condition (\ref{Eq:IntCond}) reduces to the requirement that $J_{bcd}$
be divergence-free.

Finally, the Bianchi equations (\ref{Eq:BianchiEq}) can be coupled
either to equations for metric components and Christoffel symbols, or
to equations for tetrad fields and connection coefficients, giving the
full nonlinear vacuum Einstein equations
\cite{EW64,HF96,ERW97,hFgN99,friedrich00}.

\subsection{$3+1$ split}

We assume there exists a globally defined time function $t: M \to
\Real$ such that $M$ is foliated by spacelike hypersurfaces
$\Sigma_\tau = \{ p \in M : t(p) = \tau \}$. Let $n_a =
-\alpha\nabla_a t$ be the future-directed unit normal to these slices,
where the time orientation is chosen so that the lapse function,
$\alpha$, is strictly positive. The three-metric $h_{ab}$ and
extrinsic curvature $k_{ab}$ are defined as\footnote{Many authors use
a different sign convention for $k_{ab}$. Our convention is that
positive mean curvature implies positive expansion of the volume
element associated with $h_{ab}$ in the direction of $n_a$.}
\begin{displaymath}
h_{ab} \equiv g_{ab} + n_a n_b\; , \qquad
k_{ab} \equiv \nabla_a n_b + n_a a_b\; ,
\end{displaymath}
where $a_b \equiv n^a\nabla_a n_b$ is the acceleration
along the integral curves of $n_a$. For a one-form $v_a$ tangential to
$\Sigma_\tau$ in the sense that $v_a n^a = 0$, the spatial covariant
derivative $D_a v_b$ is defined as $D_a v_b \equiv h_a{}^c
h_b{}^d\nabla_c v_d$. The spatial covariant derivative of a general
tangential tensor field is defined similarly. The electric and
magnetic parts of $C_{abcd}$ are, respectively,
\begin{displaymath}
E_{ab} = C_{acbd} n^c n^d, \qquad
H_{ab} = \frac{1}{2} n^c C_{caef} \varepsilon^{ef}{}_b\; ,
\end{displaymath}
where $\varepsilon_{bcd} = n^a\varepsilon_{abcd}$ denotes the natural
volume element on $(\Sigma_t,h_{ab})$. From the symmetries
(\ref{Eq:WeylSym}) of $C_{abcd}$, it follows that $E_{ab}$ and
$H_{ab}$ are symmetric, traceless, and orthogonal to
$n^a$. Furthermore, the ten fields $\{E_{ab}$, $H_{ab}\}$ uniquely
determine $C_{abcd}$:
\begin{displaymath}
C_{abcd} = -4 n_{[a} E_{b][c} n_{d]} 
 - \varepsilon_{ab}{}^{e} E_{ef} \varepsilon^{f}{}_{cd}
 - 2n_{[a} H_{b]e} \varepsilon^{e}{}_{cd} 
 + 2\varepsilon_{ab}{}^{e} H_{e[c} n_{d]}\, .
\end{displaymath}
The decomposition of Eq. (\ref{Eq:BianchiEq}) into components normal
and tangential to $n^a$ yields the evolution equations
\begin{eqnarray}
\pounds_n E_{ab} &=& -\varepsilon_{cd(a} (D^c + 2a^c) H^d{}_{b)}
 + 5 k_{(a}{}^d E_{b)d} - 2 k E_{ab} 
  - h_{ab} k^{cd} E_{cd} + R_{ab}\; ,
\label{Eq:EvolWeylE}\\
\pounds_n H_{ab} &=& +\varepsilon_{cd(a} (D^c + 2a^c) E^d{}_{b)}
 + 5 k_{(a}{}^d H_{b)d} - 2 k H_{ab} 
  - h_{ab} k^{cd} H_{cd} + S_{ab}\; ,
\label{Eq:EvolWeylH}
\end{eqnarray}
and the constraint equations
\begin{eqnarray}
D^b E_{ab} - k^{cd}\varepsilon^b{}_{da} H_{cb} &=& P_a\; ,
\label{Eq:DivE}\\
D^b H_{ab} + k^{cd}\varepsilon^b{}_{da} E_{cb} &=& Q_a\; .
\label{Eq:DivH}
\end{eqnarray}
In these equations, $k \equiv h^{ab} k_{ab}$,
\begin{eqnarray}
   P_c = n^b n^d J_{bcd}\, , \qquad
&& Q_a = -\frac{1}{2}\, n^b \varepsilon_{a}{}^{cd} J_{bcd}\, ,
\nonumber\\
   R_{ef} = -h_{(e}{}^{b} n^c h_{f)}{}^{d} J_{bcd}\, ,\qquad
&& S_{ef} = -\frac{1}{2}\, h_{(e}{}^{b} \varepsilon_{f)}{}^{cd} J_{bcd}\, ,
\nonumber
\end{eqnarray}
and $\pounds_n$ denotes the Lie derivative with respect to the unit
normal field $n^a$. Notice that
Eqs. (\ref{Eq:EvolWeylE},\ref{Eq:EvolWeylH}) and
(\ref{Eq:DivE},\ref{Eq:DivH}) obey the ``Dirac duality'' symmetry
\begin{equation}
(E_{ab},H_{ab}) \mapsto (H_{ab},-E_{ab}),\qquad 
(P_a,Q_a) \mapsto (Q_a,-P_a),\qquad
(R_{ab},S_{ab}) \mapsto (S_{ab},-R_{ab}).
\label{Eq:DiracDuality}
\end{equation}

\subsection{$2+1$ split}
\label{SubSect:2+1}

In addition to the foliation $\Sigma_\tau$ by spacelike hypersurfaces,
the existence of a unit spatial vector field $s^a$ which is everywhere
tangential to the hypersurfaces $\Sigma_t$ is assumed. The existence
of such a vector field allows us to introduce a Newman-Penrose null
tetrad
\begin{displaymath}
l^a = \frac{1}{\sqrt{2}}\left( n^a + s^a \right), \qquad
k^a = \frac{1}{\sqrt{2}}\left( n^a - s^a \right), \qquad
m^a = \frac{1}{\sqrt{2}}\left( v^a + i\, w^a \right), \qquad
\bar{m}^a = \frac{1}{\sqrt{2}}\left( v^a - i\, w^a \right),
\end{displaymath}
where $v^a$ and $w^a$ are two mutually orthogonal unit vector fields
which are normal to $n^a$ and $s^a$. The corresponding Newman-Penrose
Weyl scalars \cite{eNrP62} are defined as\footnote{Notice that we use
a different sign convention for the metric and for $\Psi_3$ than in
Ref. \cite{eNrP62}.}
\begin{eqnarray}
\Psi_0 &=& C_{abcd} l^a m^b l^c m^d,
\nonumber\\
\Psi_1 &=& C_{abcd} l^a k^b l^c m^d,
\nonumber\\
\Psi_2 &=& C_{abcd} l^a m^b \bar{m}^c k^d,
\nonumber\\
\Psi_3 &=& C_{abcd} l^a k^b k^c\bar{m}^d,
\nonumber\\
\Psi_4 &=& C_{abcd} k^a \bar{m}^b k^c \bar{m}^d.
\nonumber
\end{eqnarray}

Next, we decompose $E_{ab}$ and $H_{ab}$ into components parallel and
normal to $s_a$. More precisely, we write
\begin{eqnarray}
E_{ab} &=& \left( s_a s_b - \frac{1}{2}\gamma_{ab} \right)\bar{E}
        + 2 s_{(a} \bar{E}_{b)} + \hat{E}_{ab}\; ,
\nonumber\\
H_{ab} &=& \left( s_a s_b - \frac{1}{2}\gamma_{ab} \right)\bar{H}
        + 2 s_{(a} \bar{H}_{b)} + \hat{H}_{ab}\; ,
\nonumber
\end{eqnarray}
where $\gamma_{ab} = h_{ab} - s_a s_b$, $\bar{E} = E_{ab} s^a s^b$,
$\bar{E}_a = \gamma_a{}^b E_{bc} s^c$, and $\hat{E}_{ab} =
\left(\gamma_a{}^c\gamma_b{}^d - \frac{1}{2}\gamma_{ab}\gamma^{cd}
\right)E_{cd}$ (with similar expressions for $\bar{H}$, $\bar{H}_a$
and $\hat{H}_{ab}$). In terms of these quantities, the Weyl scalars
are
\begin{eqnarray}
\Psi_0 &=& \left[ \hat{E}_{ab} + \varepsilon_a{}^c\hat{H}_{cb} \right] m^a m^b,
\label{Eq:Psi0}\\
\Psi_1 &=& 
 -\frac{1}{\sqrt{2}} \left[ \bar{E}_a + \varepsilon_a{}^b\bar{H}_b \right] m^a,
\label{Eq:Psi1}\\
\Psi_2 &=& \frac{1}{2} \left[ \bar{E} - i\bar{H} \right],
\label{Eq:Psi2}\\
\Psi_3 &=& 
 -\frac{1}{\sqrt{2}} \left[ \bar{E}_a - \varepsilon_a{}^b\bar{H}_b \right]
 \bar{m}^a,
\label{Eq:Psi3}\\
\Psi_4 &=& \left[ \hat{E}_{ab} - \varepsilon_a{}^c\hat{H}_{cb} \right] 
 \bar{m}^a \bar{m}^b,
\label{Eq:Psi4}
\end{eqnarray}
where $\varepsilon_{ab} = \varepsilon_{abc} s^c$. 

In order to decompose the evolution equations
(\ref{Eq:EvolWeylE},\ref{Eq:EvolWeylH}), we make additional
assumptions on the vector fields $n^a$ and $s^a$. First, we assume
that $s^a$ is geodetic and everywhere orthogonal to closed
$2$-surfaces $S_r$ in $\Sigma_t$. This implies that
\begin{displaymath}
D_a s_b = \kappa_{ab}\; ,
\end{displaymath}
where $\kappa_{ab}$ is a symmetric tensor field which is orthogonal to
$s^a$, representing the extrinsic curvature of the $2$-surfaces $S_r$
as embedded in $\Sigma_t$. Next, we assume that the Lie-derivative
$\pounds_n s^a$ of the vector field $s^a$ with respect to $n^a$ can be
written as a linear combination of $n^a$ and $s^a$. This implies that
any covariant tensor field $t_{a_1 a_2 ... a_k}$ which is orthogonal
to $n^a$ and $s^a$ has the property that $\pounds_n t_{a_1 a_2
... a_k}$ is again orthogonal to $n^a$ and $s^a$. Finally, we assume
that the Lie-derivative $\pounds_n s_a$ of the one-form $s_a$ is
proportional to $s_a$. These properties, together with the relations
$n_a s^a = 0$, $s_a s^a = 1$, $\pounds_n n_a = D_a(\log\alpha)$, and
$\pounds_n h_{ab} = 2k_{ab}$, imply that
\begin{displaymath}
\bar{k}_a \equiv \gamma_a{}^b k_{bc} s^c = 0, \qquad
\pounds_n s^a = (\pounds_s\log\alpha) n^a -\bar{k} s^a, \qquad
\pounds_n s_a = \bar{k} s_a\; ,
\end{displaymath}
where $\bar{k} = k_{ab} s^a s^b$. Although these assumptions are
strong, and may not all hold for a generic spacetime, they are
satisfied for the background spacetimes and foliations used in this
article. In particular, they are met for any spherically symmetric
spacetime of the form
\begin{displaymath}
ds^2 = -\alpha^2 dt^2 + \gamma^2\left( dr + \beta\, dt \right)^2
 + r^2\left( d\vartheta^2 + \sin^2\vartheta\, d\varphi^2 \right),
\end{displaymath}
where $\alpha$, $\beta$ and $\gamma$ are smooth functions of $t$ and
$r$, and where $n_a dx^a = -\alpha dt$ and $s_a dx^a = \gamma(dr +
\beta dt)$. Decomposing the extrinsic curvature $k_{ab}$ as
\begin{displaymath}
k_{ab} = \left( s_a s_b - \frac{1}{2}\gamma_{ab} \right)\bar{k} 
       + \hat{k}_{ab} + \frac{1}{2}\gamma_{ab} k,
\end{displaymath}
where $\hat{k}_{ab} = \left(\gamma_a{}^c\gamma_b{}^d -
\frac{1}{2}\gamma_{ab}\gamma^{cd} \right)k_{cd}$, we find from
evolution equation (\ref{Eq:EvolWeylE}) that
\begin{eqnarray}
\pounds_n \bar{E} &=& -\frac{1}{\alpha^2}
   \varepsilon^{ab}{\cal D}_a\left( \alpha^2 \bar{H}_b \right) 
 + \hat{\kappa}^{ab} \varepsilon_a{}^c\hat{H}_{cb}
 + \frac{3}{2} \left( \bar{k} - k \right)\bar{E} 
 - \hat{k}^{ab}\hat{E}_{ab} + \bar{R}\; ,
\label{Eq:Ebar}\\
\pounds_n \bar{E}_a &=& 
  \frac{1}{2\alpha^2}\pounds_s\left( \alpha^2\varepsilon_a{}^b\bar{H}_b \right)
 -\frac{1}{2\alpha^2}{\cal D}^c\left( \alpha^2\varepsilon_a{}^d\hat{H}_{cd} \right)
 -\frac{3}{4\alpha^2}\varepsilon_a{}^b{\cal D}_b\left( \alpha^2 \bar{H} \right)
\nonumber\\
 &-& 2\hat{\kappa}_a{}^b\varepsilon_b{}^c\bar{H}_c
  + \frac{1}{4}\left( \bar{k} - 3k \right)\bar{E}_a
  + \frac{5}{2}\hat{k}_a{}^b\bar{E}_b + \bar{R}_a\; ,
\label{Eq:Ebara}\\
\pounds_n \hat{E}_{ab} &=&
 \left[ \frac{1}{\alpha^2}
        \pounds_s\left( \alpha^2\varepsilon_a{}^c\hat{H}_{bc} \right)
 + \frac{1}{\alpha^2}
        \varepsilon_{c(a} {\cal D}^c\left( \alpha^2 \bar{H}_{b)} \right)
 - \left( 3\hat{\kappa}_{c(a} + \frac{1}{2}\gamma_{c(a}\kappa \right)
   \varepsilon^{cd}\hat{H}_{b)d}
 - \frac{3}{2} \varepsilon_{(a}{}^c \hat{\kappa}_{b)c}\bar{H}
 \right]^{tf}
\nonumber\\
 &-& \frac{1}{2}\left( 5\bar{k} - k \right)\hat{E}_{ab}
  - \frac{3}{2}\hat{k}_{ab}\bar{E}
  + 5\hat{k}_{(a}{}^c\hat{E}_{b)c}
  - \frac{3}{2}\gamma_{ab}\hat{k}^{cd}\hat{E}_{cd} + \hat{R}_{ab}\; ,
\label{Eq:Ebarab}
\end{eqnarray}
where $[...]^{tf}$ denotes the trace-free part with respect to
$\gamma_{ab}$, $\{\kappa$, $\hat{\kappa}_{ab}\}$ denote the trace and
trace-free part of $\kappa_{ab}$, respectively, ${\cal D}$ denotes the
covariant derivative compatible with $\gamma_{ab}$, and $\{\bar{R}$,
$\bar{R}_a$, $\hat{R}_{ab}\}$ denote the parallel/parallel,
parallel/transverse, and transverse trace-free parts of $R_{ab}$,
respectively. The constraint equation (\ref{Eq:DivE}) yields
\begin{eqnarray}
\bar{P} &=& \pounds_s \bar{E} + {\cal D}^a\bar{E}_a 
         - \hat{\kappa}^{ab}\hat{E}_{ab} + \frac{3\kappa}{2}\bar{E}
         + \hat{k}^{ab}\varepsilon_a{}^c\hat{H}_{bc}\; ,
\label{Eq:PbarDef}\\
\bar{P}_a &=& \pounds_s \bar{E}_a + {\cal D}^b\hat{E}_{ab}
  - \frac{1}{2} {\cal D}_a\bar{E} + \kappa\bar{E}_a
  - \frac{1}{2}\left( 3\bar{k} - k \right)\varepsilon_a{}^b\bar{H}_b
  - \hat{k}_a{}^b\varepsilon_b{}^c\bar{H}_c\; ,
\label{Eq:PbaraDef}
\end{eqnarray}
where we have defined $\bar{P} = P_a s^a$ and $\bar{P}_a =
\gamma_a{}^b P_b$. Evolution and constraint equations for $\bar{H}$,
$\bar{H}_a$ and $\hat{H}_{ab}$ are easily obtained by applying
the Dirac duality transformations (\ref{Eq:DiracDuality}).

\subsection{Propagation of the constraint fields}

The decomposition of the integrability condition (\ref{Eq:IntCond})
into parts normal and tangential to $n^a$ gives (assuming that the
background metric is flat or conformally flat)
\begin{eqnarray}
\pounds_n P_a &=& -\frac{1}{2} \varepsilon_a^{\;\; cd} (D_c + 3a_c) Q_d 
 + \frac{3}{2} \left( k_{a}{}^{b} P_b - k P_a \right)
 + (D^b + a^b) R_{ab} + \varepsilon_{a}{}^{cd} k^{b}{}_{c} S_{bd}\, ,
\label{Eq:Pa}\\
\pounds_n Q_a &=& \frac{1}{2} \varepsilon_a^{\;\; cd} (D_c + 3a_c) P_d 
 + \frac{3}{2} \left( k_{a}{}^{b} Q_b - k Q_a \right)
 + (D^b + a^b) S_{ab} - \varepsilon_{a}{}^{cd} k^{b}{}_{c} R_{bd}\, .
\label{Eq:Qa}
\end{eqnarray}
If the evolution equations (\ref{Eq:EvolWeylE},\ref{Eq:EvolWeylH})
hold with $R_{ab} = S_{ab} = 0$, then the fields $P_a$ and $Q_a$ obey
homogeneous Maxwell-like equations with transmission speed half the
speed of light. In particular, $P_a = Q_a = 0$ on an initial spatial
slice gives $P_a = Q_a = 0$ on the future domain of dependence of the
initial slice. We may therefore regard $P_a = Q_a = 0$ as constraints
which are propagated by the evolution equations
(\ref{Eq:EvolWeylE},\ref{Eq:EvolWeylH}) with $R_{ab} = S_{ab} = 0$.

Under the same assumptions on the vector fields $n^a$ and $s^a$ as in
the previous subsection, the $2+1$ split of the constraint propagation
equations yields
\begin{eqnarray}
\pounds_n \bar{P} &=& 
 -\frac{1}{2\alpha^3}\varepsilon^{ab}{\cal D}_a\left( \alpha^3\bar{Q}_b\right)
 + \frac{1}{2}\left(\bar{k} - 3k \right)\bar{P}
\nonumber\\
 &+& \frac{1}{\alpha}\pounds_s(\alpha\bar{R})
  + \frac{1}{\alpha}{\cal D}^b(\alpha\bar{R}_b)
  - \hat{\kappa}^{ab}\hat{R}_{ab} + \frac{3\kappa}{2}\bar{R}
  + \hat{k}^{ab}\varepsilon_a{}^c\hat{S}_{bc}\; ,
\label{Eq:Pbar}\\
\pounds_n \bar{P}_a &=& \frac{1}{2\alpha^3}\varepsilon_a{}^b
 \left[ \pounds_s(\alpha^3\bar{Q}_b) - {\cal D}_b(\alpha^3\bar{Q}) \right]
 - \frac{3}{4}\left( \bar{k} + k \right)\bar{P}_a 
 + \frac{3}{2}\hat{k}_a{}^b\bar{P}_b
\nonumber\\
 &+& \frac{1}{\alpha}\pounds_s(\alpha\bar{R}_a)
  + \frac{1}{\alpha}{\cal D}^b(\alpha\hat{R}_{ab})
  - \frac{1}{2\alpha}{\cal D}_a(\alpha\bar{R})
  + \kappa\bar{R}_a 
  - \frac{1}{2}\left( 3\bar{k} - k \right)\varepsilon_a{}^b\bar{S}_b
  - \hat{k}_a{}^b\varepsilon_b{}^c\bar{S}_c\; .
\label{Eq:Pbara}
\end{eqnarray}
The corresponding equations for $\bar{Q}$ and $\bar{Q}_a$ are
obtained from this by applying the Dirac duality transformations
(\ref{Eq:DiracDuality}).

\section{Exact solutions on a Minkowski background}
\label{Sect:Exact}

In this section, we consider the Bianchi equations
(\ref{Eq:BianchiEq}) on the Minkowski background
\begin{displaymath}
ds^2 = -dt^2 + dr^2 + r^2 \hat{g}_{AB} dx^A dx^B,
\end{displaymath}
where $\hat{g}_{AB} dx^A dx^B = d\vartheta^2 + \sin^2\vartheta\;
d\varphi^2$ denotes the standard metric on $S^2$. A natural foliation
is given by the slices $t=const.$, for which $n_a dx^a = -dt$,
although other foliations are possible. (In particular, it should be
interesting to generalize the investigation below to hyperboloidal
slices.) Since the spacetime is spherically symmetric, it is natural
to choose $s_a dx^a = dr$. The corresponding vector field $s^a$ is
defined everywhere except at the center $r=0$. Furthermore, $\kappa =
2/r$ and $\hat{\kappa}_{ab} = 0$, so the evolution and constraint
equations derived in the previous section simplify considerably. In
this section, we assume that the source terms $R_{ab}$ and $S_{ab}$
vanish identically; however, we do not necessarily enforce the
constraints $P_a = Q_a = 0$ since we are also interested in studying
the propagation of constraint violations.

\subsection{Harmonic decomposition}

Since the spacetime is spherically symmetric and the equations are
linear, it is convenient to expand the fields in spherical tensor
harmonics. In the resulting equations, pieces belonging to different
angular momentum numbers $\ell$ and $m$ decouple. Thus, it is
sufficient to consider one fixed value of $\ell$ and $m$ at a time.
The decomposition of the fields $E_{ab}$ and $H_{ab}$ into spherical
tensor harmonics reads
\begin{eqnarray}
\bar{E} &=& \frac{1}{r} e_0(t,r) Y, 
\nonumber\\
\bar{E}_A &=& e_1(t,r) \hat{\nabla}_A Y + f_1(t,r)\hat{S}_A\; ,
\nonumber\\
\hat{E}_{AB} &=& 2r e_2(t,r)\left[ \hat{\nabla}_A\hat{\nabla}_B \right]^{tf} Y
 + 2r f_2(t,r) \hat{\nabla}_{(A}\hat{S}_{B)}\; ,
\label{Eq:EHHarmDecomp}\\
\bar{H} &=& \frac{1}{r} h_0(t,r) Y, 
\nonumber\\
\bar{H}_A &=& h_1(t,r) \hat{\nabla}_A Y + g_1(t,r)\hat{S}_A\; ,
\nonumber\\
\hat{H}_{AB} &=& 2r h_2(t,r)\left[ \hat{\nabla}_A\hat{\nabla}_B \right]^{tf} Y
 + 2r g_2(t,r) \hat{\nabla}_{(A}\hat{S}_{B)}\; ,
\nonumber
\end{eqnarray}
where $Y = Y^{\ell m}(\vartheta,\varphi)$ denotes the standard
spherical harmonics, $\hat{S}_A = \varepsilon_A{}^B\hat{\nabla}_B Y$,
and $\hat{\nabla}$ denotes the covariant derivative on $S^2$. Similarly,
the constraint variables can be written as
\begin{eqnarray}
&& \bar{P} = \frac{1}{r} P_0(t,r) Y, \qquad
   \bar{P}_A = P_1(t,r) \hat{\nabla}_A Y + P_2(t,r)\hat{S}_A\; ,
\label{Eq:PHarmDecomp}\\
&& \bar{Q} = \frac{1}{r} Q_0(t,r) Y, \qquad
   \bar{Q}_A = Q_1(t,r) \hat{\nabla}_A Y + Q_2(t,r)\hat{S}_A\; .
\label{Eq:QHarmDecomp}
\end{eqnarray}
The Newman-Penrose scalars are
\begin{eqnarray}
\Psi_0 &=& \frac{2}{r}(e_2 - g_2)\hat{m}^A\hat{m}^B
 \hat{\nabla}_A\hat{\nabla}_B Y
 + \frac{2}{r}(f_2 + h_2)\hat{m}^A\hat{m}^B\hat{\nabla}_A\hat{S}_B\; ,
\label{Eq:Psi0Harmonic}\\
\Psi_1 &=& -\frac{1}{\sqrt{2} r}(e_1 - g_1)\hat{m}^A\hat{\nabla}_A Y
           -\frac{1}{\sqrt{2} r}(f_1 + h_1)\hat{m}^A\hat{S}_A\; ,
\label{Eq:Psi1Harmonic}\\
\Psi_2 &=& \frac{1}{2r}\left( e_0 -i h_0 \right) Y,
\label{Eq:Psi2Harmonic}\\
\Psi_3 &=& -\frac{1}{\sqrt{2} r}(e_1 + g_1)\bar{\hat{m}}^A\hat{\nabla}_A Y
           -\frac{1}{\sqrt{2} r}(f_1 - h_1)\bar{\hat{m}}^A\hat{S}_A\; ,
\label{Eq:Psi3Harmonic}\\
\Psi_4 &=& \frac{2}{r}(e_2 + g_2)\bar{\hat{m}}^A\bar{\hat{m}}^B
 \hat{\nabla}_A\hat{\nabla}_B Y
 + \frac{2}{r}(f_2 - h_2)\bar{\hat{m}}^A\bar{\hat{m}}^B
   \hat{\nabla}_A\hat{S}_B\; ,
\label{Eq:Psi4Harmonic}
\end{eqnarray}
where $\hat{m}^A = r\, m^A$. Using the identities
\begin{eqnarray}
\hat{\nabla}^B \left[ \hat{\nabla}_A\hat{\nabla}_B \right]^{tf} Y
 &=& -\frac{\lambda}{2} \hat{\nabla}_A Y, \\
\hat{\nabla}^B \hat{\nabla}_{(A}\hat{S}_{B)} 
 &=& -\frac{\lambda}{2} \hat{S}_A\; , 
\end{eqnarray}
where $\lambda = (\ell-1)(\ell+2)$, the Bianchi equations yield a set
of two decoupled systems for the amplitudes $(e_0,e_1,e_2,g_1,g_2)$
(even parity sector) and $(h_0,h_1,h_2,f_1,f_2)$ (odd-parity sector).
For $\ell\geq 2$, the even parity sector is described by the evolution
system
\begin{eqnarray}
\dot{e}_0 &=& -\frac{\ell(\ell+1)}{r} g_1\; ,
\label{Eq:e0}\\
\dot{e}_1 &=& -\frac{1}{2} g_1' - \frac{\lambda}{2r} g_2\; ,
\label{Eq:e1}\\
\dot{e}_2 &=& -g_2' + \frac{1}{2r} g_1\; ,
\label{Eq:e2}\\
\dot{g}_1 &=& -\frac{1}{2} e_1' - \frac{\lambda}{2r} e_2 + \frac{3}{4r} e_0\; ,
\label{Eq:g1}\\
\dot{g}_2 &=& -e_2' + \frac{1}{2r} e_1\; ,
\label{Eq:g2}
\end{eqnarray}
where here and in the following, a dot and a prime denote
differentiation with respect to $t$ and $r$, respectively. The
evolution system
(\ref{Eq:e0},\ref{Eq:e1},\ref{Eq:e2},\ref{Eq:g1},\ref{Eq:g2}) is
subject to the constraints $P_0 = P_1 = Q_2 = 0$, where
\begin{eqnarray}
P_0 &=& \frac{1}{r^2} (r^2 e_0)' - \frac{\ell(\ell+1)}{r} e_1 \; ,
\label{Eq:P0Def}\\
P_1 &=& \frac{1}{r^2} (r^2 e_1)' - \frac{\lambda}{r} e_2 - \frac{1}{2r} e_0\; ,
\label{Eq:P1Def}\\
Q_2 &=& \frac{1}{r^2} (r^2 g_1)' - \frac{\lambda}{r} g_2\; .
\label{Eq:Q2Def}
\end{eqnarray}
The odd-parity sector is obtained from this after the substitutions
$(e_0,e_1,e_2,g_1,g_2) \mapsto (h_0,h_1,h_2,-f_1,-f_2)$. Therefore, it
is sufficient to discuss the even parity sector, which is what we will
do in the following.

\subsection{Exact solutions}
\label{SubSect:ExactSolutions}

In this section, we discuss how to obtain exact analytic solutions to
this constrained evolution system. We start with the special cases
$\ell=0$ and $\ell=1$ which, as we will see, are non-radiative. For
$\ell=0$, the only equations are $\dot{e}_0 = 0$ and $(r^2 e_0)' =
0$ which yield the solution
\begin{equation}
e_0 = -\frac{2M}{r^2}\; , \qquad M = const.
\label{Eq:SchwarzschildMode}
\end{equation}
For $\ell=1$, $\hat{E}_{AB}$ and $\hat{H}_{AB}$ vanish, and the
evolution equations for $e_2$ and $g_2$, and the constraint equation
$Q_2 = 0$, are void. Taking a time derivative of the constraint $P_0 =
0$, and eliminating $\dot{e}_0$ and $\dot{e}_1$ using the evolution
equations, one obtains $g_1 = c(t)/r^2$, where the function $c$ is
independent of $r$. The insertion of this information back into the
evolution equations for $e_0$ and $e_1$ gives $e_0 = -2e_1 + k(r)$ for
a function $k$ which is independent of $t$. Substitution of these
results into the evolution equation for $g_1$ and the constraint $P_1
= 0$ gives $c(t) = c_2 t + c_1$, $k(r) = 2c_2/r$ for some constants
$c_1$ and $c_2$, and $e_1 = c_2/(2r) + f(t)/r^3$ for a function $f$
which is independent of $r$. Finally, the insertion of all this into
the evolution equation for $e_0$ yields $\dot{f} = c_2 t + c_1$. Thus,
one obtains the most general solution in the sector $\ell=1$:
\begin{displaymath}
e_0 = \frac{c_2}{r} - \frac{c_2 t^2 + 2c_1 t + 2c_0}{r^3}\; ,\qquad
e_1 = \frac{c_2}{2r} + \frac{c_2 t^2 + 2c_1 t + 2c_0}{2r^3}\; ,\qquad
g_1 = \frac{c_2 t + c_1}{r^2}\; .
\end{displaymath}
If we demand that the solution be stationary, then the corresponding
odd-parity solution is
\begin{equation}
h_0 = \frac{6J}{r^3}\; , \qquad
h_1 = -\frac{3J}{r^3}\; , \qquad
f_1 = 0, \qquad  J = const.
\label{Eq:KerrMode}
\end{equation}
With the normalization $Y^{(\ell=0)} = 1$, $Y^{(\ell=1)} =
\cos\vartheta$, Eqs. (\ref{Eq:SchwarzschildMode},\ref{Eq:KerrMode})
yield the linearized Weyl tensor belonging to the linearized Kerr
metric
\begin{displaymath}
 -dt^2 + dr^2 + r^2\left( d\vartheta^2 + \sin^2\vartheta\, d\varphi^2 \right)
  + \frac{2M}{r}\left( dt^2 + dr^2 \right) 
  - \frac{4J}{r}\, \sin^2\vartheta\, d\varphi dt,
\end{displaymath}
where the linearization is performed about flat spacetime in the mass
parameter $M$ and the angular momentum parameter $J$.

Next, we consider the cases with $\ell\geq 2$. We obtain two
classes of solutions, one representing outgoing radiation and the
other representing incoming radiation. If the constraints are
satisfied, these solutions correspond to those presented in
\cite{wB71,aAcE88,Teukolsky82}. To solve the above constrained
evolution system, we first derive the constraint propagation system,
which describes the propagation of constraint violations under the
flux defined by the evolution equations
(\ref{Eq:EvolWeylE},\ref{Eq:EvolWeylH}). The constraint propagation
system can be obtained either by performing a multipolar decomposition
of the evolution system (\ref{Eq:Pbar},\ref{Eq:Pbara}) with $R_{ab} =
S_{ab} = 0$, or by taking a time derivative of
Eqs. (\ref{Eq:P0Def},\ref{Eq:P1Def},\ref{Eq:Q2Def}) and using the
evolution equations
(\ref{Eq:e0},\ref{Eq:e1},\ref{Eq:e2},\ref{Eq:g1},\ref{Eq:g2}). The
result is
\begin{eqnarray}
\dot{P}_0 &=& -\frac{\ell(\ell+1)}{2r}\, Q_2\; ,
\label{Eq:P0}\\
\dot{P}_1 &=& -\frac{1}{2}\, Q_2'\; ,
\label{Eq:P1}\\
\dot{Q}_2 &=& -\frac{1}{2}\left( P_1' - \frac{1}{r} P_0 \right).
\label{Eq:Q2}
\end{eqnarray}
Solutions to this system have the form
\begin{equation}
P_0 = -\frac{\ell(\ell+1)}{2r}\, \pi + r h'(r), \qquad
P_1 = -\frac{1}{2}\pi' + h(r), \qquad
Q_2 = \dot{\pi},
\label{Eq:P0P1Q2Sol}
\end{equation}
where $h(r)$ is a function of $r$ only and where the function $\pi$
satisfies the master equation
\begin{equation}
\left[ \frac{1}{c^2}\partial_t^2 - \partial_r^2 
 + \frac{\ell(\ell+1)}{r^2} \right]\pi(t,r) = 0,
\label{Eq:MasterConstraintEvol}
\end{equation}
with $c=1/2$. Eq. (\ref{Eq:MasterConstraintEvol}) describes the
evolution of constraint violations which propagate at half the speed
of light. How to solve Eq. (\ref{Eq:MasterConstraintEvol}) will be
explained below.

Once the constraint variables $P_0$, $P_1$ and $Q_2$ have been
obtained, we proceed as follows. First, using
Eqs. (\ref{Eq:P0Def},\ref{Eq:P1Def},\ref{Eq:Q2Def}) we express
$e_1$, $e_2$ and $g_2$ in terms of $e_0$ and $g_1$ and the constraint
variables:
\begin{eqnarray}
\ell(\ell+1) e_1 &=& \frac{1}{r}\phi' - r P_0\; ,
\label{Eq:e1solved}\\
\lambda e_2 &=& \frac{1}{\ell(\ell+1) r} \left[ r\phi' - r^3 P_0 \right]'
 - \frac{1}{2r^2}\phi - r P_1\; ,
\label{Eq:e2solved}\\
\lambda g_2 &=& \frac{1}{r}( r^2 g_1 )' - r Q_2\; ,
\label{Eq:g2solved}
\end{eqnarray}
where we have set $\phi = r^2 e_0$. Next, using these expressions in
Eqs. (\ref{Eq:e0}) and (\ref{Eq:g1}), we obtain the following wave
equation for $\phi$:
\begin{equation}
\left[ \partial_t^2 - \partial_r^2 
 + \frac{\ell(\ell+1)}{r^2} \right]\phi(t,r) = S(t,r),
\label{Eq:MasterEvol}
\end{equation}
where the source term $S(t,r)$ depends on the constraint variables
$\pi(t,r)$ and $h(r)$ and is given by
\begin{displaymath}
S(t,r) = \frac{\ell(\ell+1)}{4}\left[ 3r\pi' + 2\pi - 2r h(r) \right] 
       - [r^3 h'(r)]'. 
\end{displaymath}
Once Eq. (\ref{Eq:MasterEvol}) has been solved for $\phi(t,r)$, the
quantities $e_1$, $e_2$ and $g_2$ are obtained from
Eqs. (\ref{Eq:e1solved},\ref{Eq:e2solved},\ref{Eq:g2solved}). Therefore,
the linearized equations reduce to the two master equations
(\ref{Eq:MasterConstraintEvol},\ref{Eq:MasterEvol}). 

We now discuss how to obtain exact solutions to these equations. We
start with the homogeneous case where $S(t,r) = 0$. For the following,
it is convenient to introduce for each $\ell = 0,1,2,...$ the
operators \cite{wB71}
\begin{displaymath}
a_\ell \equiv \partial_r + \frac{\ell}{r} 
            = r^{-\ell}\partial_r( r^{\ell} . ),
\end{displaymath}
and their formal adjoints
\begin{displaymath}
a_\ell^\dagger \equiv -\partial_r + \frac{\ell}{r} 
                   = -r^\ell\partial_r( r^{-\ell} . ).
\end{displaymath}
They satisfy the operator identities
\begin{equation}
a_{\ell+1} a_{\ell+1}^\dagger = a_\ell^\dagger a_\ell 
 = -\partial_r^2 + \frac{\ell(\ell+1)}{r^2}\; .
\label{Eq:ComRel}
\end{equation}
As a consequence, for each $\ell=0,1,2,...$,
\begin{eqnarray}
\left[ \partial_t^2 - \partial_r^2 + \frac{\ell(\ell+1)}{r^2} \right]
a_\ell^\dagger a_{\ell-1}^\dagger ... a_1^\dagger
 &=& \left[ \partial_t^2 + a_{\ell}^\dagger a_{\ell} \right]
   a_\ell^\dagger a_{\ell-1}^\dagger ... a_1^\dagger
\nonumber\\
 &=& a_\ell^\dagger \left[ \partial_t^2 + a_{\ell-1}^\dagger a_{\ell-1} \right]
     a_{\ell-1}^\dagger ... a_1^\dagger
\nonumber\\
 &=& a_\ell^\dagger a_{\ell-1}^\dagger ... a_1^\dagger
     \left[ \partial_t^2 - \partial_r^2 \right].
\label{Eq:ComRelbis}
\end{eqnarray}
Therefore, in- and outgoing solutions to the homogeneous master
equation can be constructed from in- and outgoing solutions of the
one-dimensional wave equation \cite{wB71}. For $\ell=0$, the in- and
outgoing solutions are given, respectively, by $\phi_{\nwarrow,0}(t,r)
\equiv V_0(r+t)$ and $\phi_{\nearrow,0}(t,r) \equiv U_0(r-t)$, where
$V_0$ and $U_0$ are smooth functions. For $\ell > 0$ the solutions
have the form
\begin{eqnarray}
\phi_{\nwarrow,\ell}(t,r) &=& a_\ell^\dagger a_{\ell-1}^\dagger ... a_1^\dagger
V_\ell(r+t),
\nonumber\\
\phi_{\nearrow,\ell}(t,r) &=& a_\ell^\dagger a_{\ell-1}^\dagger ... a_1^\dagger
U_\ell(r-t),
\nonumber
\end{eqnarray}
where $V_\ell$ and $U_\ell$ are sufficiently smooth functions.
Explicit expressions for $\phi_{\nwarrow,\ell}$ and
$\phi_{\nearrow,\ell}$ are given by
\begin{eqnarray}
\phi_{\nwarrow,\ell}(t,r) 
&=& (-1)^\ell r^{\ell}  \left( \frac{d}{dr}\frac{1}{r} \right)^\ell V_\ell(r+t)
 = \sum\limits_{j=0}^\ell (-1)^j\frac{(2\ell-j)!}{(\ell-j)!\, j!} 
     (2r)^{j-\ell} V^{(j)}_\ell(r+t),
\label{Eq:ExplicitSolFormIn}\\
\phi_{\nearrow,\ell}(t,r) 
&=& (-1)^\ell r^{\ell}  \left( \frac{d}{dr}\frac{1}{r} \right)^\ell U_\ell(r-t)
 = \sum\limits_{j=0}^\ell (-1)^j\frac{(2\ell-j)!}{(\ell-j)!\, j!} 
     (2r)^{j-\ell} U^{(j)}_\ell(r-t),
\label{Eq:ExplicitSolFormOut}
\end{eqnarray}
where here and in the following, for a function $F$ on the real line,
$F^{(j)}$ denotes its $j$'th derivative. As an example, for $\ell=2$,
\begin{displaymath}
\phi_{\nearrow,2}(t,r) = \frac{3}{r^2} U_2(r-t)
                       - \frac{3}{r}U^{(1)}_2(r-t) + U^{(2)}_2(r-t).
\end{displaymath}
In- and outgoing solutions of the constraint propagation master
equation (\ref{Eq:MasterConstraintEvol}) can be obtained in exactly
the same way after replacing $t$ by $c t$, {\em ie.},
\begin{eqnarray}
\pi_{\nwarrow,\ell}(t,r) &=& a_\ell^\dagger a_{\ell-1}^\dagger ... a_1^\dagger
W_\ell(r+ct),
\nonumber\\
\pi_{\nearrow,\ell}(t,r) &=& a_\ell^\dagger a_{\ell-1}^\dagger ... a_1^\dagger
Z_\ell(r-ct),
\nonumber\\
\end{eqnarray}
for some sufficiently smooth functions $W_\ell$ and $Z_\ell$.

Finally, we discuss the case $S(t,r) \neq 0$. Since we have already
calculated the solutions to the homogeneous problem, it is sufficient
to construct one particular solution of Eq. (\ref{Eq:MasterEvol}). In
the following, we assume that $h \equiv 0$ and that $\pi(t,r)$ has the
form $\pi(t,r) = a_\ell^\dagger a_{\ell-1}^\dagger ... a_1^\dagger
W_\ell(t,r)$, with $W_\ell$ a sufficiently smooth function of $t$ and
$r$. The latter assumption is no restriction of generality, since we
can obtain $W_\ell$ from $\pi$ by successive integration,
\begin{displaymath}
W_\ell(t,r) = r\int\limits_r^\infty dr_1 r_1 \int\limits_{r_1}^\infty dr_2 r_2 
 ... \int\limits_{r_{\ell-1}}^\infty dr_\ell \frac{\pi(t,r_\ell)}{r_\ell^\ell}
\; ,
\end{displaymath}
provided that $\pi(t,.)$ falls off sufficiently rapidly as
$r\to\infty$. To construct a particular solution $\phi_1(t,r)$ of
Eq. (\ref{Eq:MasterEvol}), we first notice that the operators
\begin{displaymath}
p_m = 3r\partial_r + m,
\end{displaymath}
$m = ...-2,-1,0,1,2,...$ satisfy the commutation relations
\begin{displaymath}
p_m a_\ell^\dagger = a_\ell^\dagger p_{m-3}\; .
\end{displaymath}
As a consequence,
\begin{displaymath}
S(t,r) = \frac{\ell(\ell+1)}{4}\, p_2 \pi(t,r) 
 = \frac{\ell(\ell+1)}{4}\, a_\ell^\dagger a_{\ell-1}^\dagger ... a_1^\dagger
 p_{2-3\ell} W_\ell(t,r).
\end{displaymath}
If we make the ansatz
\begin{displaymath}
\phi_1(t,r) = a_\ell^\dagger a_{\ell-1}^\dagger ... a_1^\dagger \psi(t,r),
\end{displaymath}
and use relation (\ref{Eq:ComRelbis}), we see that $\phi_1$ is a
particular solution if $\psi$ satisfies the inhomogeneous
one-dimensional wave equation
\begin{displaymath}
\left[ \partial_t^2 - \partial_r^2 \right] \psi(t,r) 
 = \frac{\ell(\ell+1)}{4} p_{2-3\ell} W_\ell(t,r).
\end{displaymath}
With trivial initial data, this equation has the solution
\begin{displaymath}
\psi(t,r) = \frac{\ell(\ell+1)}{8} \int\limits_0^t 
 \int\limits_{r-t+\tau}^{r+t-\tau} p_{2-3\ell} W_\ell(\tau,s) ds d\tau.
\end{displaymath}
Therefore, a particular solution of Eq. (\ref{Eq:MasterEvol}) is given
by
\begin{equation}
\phi_1(t,r) = \frac{\ell(\ell+1)}{8} 
 a_\ell^\dagger a_{\ell-1}^\dagger ... a_1^\dagger 
\int\limits_0^t 
 \int\limits_{r-t+\tau}^{r+t-\tau} p_{2-3\ell} W_\ell(\tau,s) ds d\tau.
\end{equation}

\section{Solutions of the initial-boundary value problem}
\label{Sect:IBVPSol}

In this section, we analyze the Bianchi equations in the presence of
artificial boundaries. Specifically, we solve the equations on a
tubular subspace $M = [0,T] \times B_R$ of Minkowski spacetime, where
$B_R$ is a ball of radius $R$ in Euclidean space. Our goal is to
impose boundary conditions on the timelike boundary ${\cal T} = [0,T]
\times \partial B_R$ which are perfectly absorbing in the following
sense: for given initial data which is compactly supported in $B_R$
and which represents a purely outgoing solution, the solution to the
IBVP leads to the same solution as the solution to the global problem
(without artificial boundaries). As discussed in the introduction,
this turns out to be a challenging problem, even for simpler systems
like the wave equation in more than one dimension
\cite{bEaM77,aBeT80}. The strategy here will be to impose different
boundary conditions on ${\cal T}$ which have been proposed in the
literature, and construct exact solutions of the resulting IBVP by
using the expressions derived in the previous section. With the help
of these solutions, we analyze how ``good'' the boundary conditions
are by looking at the amount of artificial reflection of constraint
violating modes and gravitational radiation. This analysis enables us
to construct different classes of boundary conditions, where each new
class yields an improvement over the old one.

We start in the next subsection with our crudest approximation for
constructing outgoing boundary conditions, which consists of using the
symmetric hyperbolic structure of the evolution equations
(\ref{Eq:EvolWeylE},\ref{Eq:EvolWeylH}) to freeze the incoming
characteristic fields to their initial values. These boundary
conditions yield a well posed IBVP. However, as we show, they
introduce constraint-violating modes into the computational domain and
therefore fail to be perfectly absorbing at a very fundamental level.

In subsection \ref{SubSect:CPBCFreezingPsi0}, first we specify CPBC
which freeze the Weyl scalar $\Psi_0$ to its initial value. This is
done either by using the method in \cite{hFgN99}, where suitable
combinations of the constraints are added to the evolution equations,
or by setting to zero the incoming constraint fields. Then, we show
that the resulting boundary conditions introduce some reflections of
(linearized) gravitational radiation. We quantify the amount of
reflection by considering outgoing waves with wave number $k$ and
computing the reflection coefficient as a function of the
dimensionless quantity $k R$.

Finally, in subsection \ref{SubSect:CPBCImproved}, we improve the
boundary conditions considered in subsection
\ref{SubSect:CPBCFreezingPsi0}. In particular, for each $L \geq 1$, we
give CPBC which are perfectly absorbing for outgoing
gravitational radiation with angular momentum number $\ell \leq L$.

\subsection{Freezing the incoming fields}
\label{SubSect:Freezing}

Let $s_a$ be the unit outward normal one-form to the boundary
$\partial B_R$. For a symmetric hyperbolic evolution system of the
form
\begin{displaymath}
{\cal A}^a(u)\frac{\partial}{\partial x^a} u = {\cal F}(u),
\end{displaymath}
with ${\cal A}^0(u) = \identy$, the characteristic speeds and fields
with respect to $s_a$ are defined, respectively, as the eigenvalues
and projections of $u$ onto the corresponding eigenspaces of the
matrix ${\cal A}^a(u) s_a$. For the evolution system
(\ref{Eq:EvolWeylE},\ref{Eq:EvolWeylH}), these are given by
\begin{eqnarray}
& \mu = -\beta^a s_a\; , & \bar{E} = 2\re\Psi_2\; ,\qquad 
                           \bar{H} = -2\im\Psi_2\; ,
\nonumber\\
& \mu = -\frac{\alpha}{2} - \beta^a s_a\; , 
& \bar{V}^{(-)}_a = \bar{E}_a + \varepsilon_a{}^b \bar{H}_b
                  = -\sqrt{2}(\Psi_1\bar{m}_a + \bar{\Psi}_1 m_a),
\nonumber\\
& \mu = +\frac{\alpha}{2} - \beta^a s_a\; ,
& \bar{V}^{(+)}_a = \bar{E}_a - \varepsilon_a{}^b \bar{H}_b
                  = -\sqrt{2}(\Psi_3 m_a + \bar{\Psi}_3\bar{m}_a),
\nonumber\\
& \mu = -\alpha - \beta^a s_a\; ,   
& \hat{V}^{(-)}_{ab} = \hat{E}_{ab} + \varepsilon_{(a}{}^c\hat{H}_{b)c}
                     = (\Psi_0\bar{m}_a\bar{m}_b + \bar{\Psi}_0 m_a m_b),
\nonumber\\
& \mu = +\alpha - \beta^a s_a\; ,
& \hat{V}^{(+)}_{ab} = \hat{E}_{ab} - \varepsilon_{(a}{}^c\hat{H}_{b)c}
                     = (\Psi_4 m_a m_b + \bar{\Psi}_4\bar{m}_a\bar{m}_b),
\nonumber
\end{eqnarray}
where we use the same notation as in Sect. \ref{SubSect:2+1}, and
where $\beta^a$ denotes the shift vector field. The ingoing fields are
the ones with negative characteristic speeds $\mu$. One way to obtain
a well posed IBVP is to freeze the ingoing fields to their initial
values \cite{pS96b}. In our choice of coordinates, with $\alpha=1$ and
$\beta^a = 0$, this boundary condition is equivalent to imposing
\begin{displaymath}
\partial_t\Psi_0 \hateq 0, \qquad
\partial_t\Psi_1 \hateq 0,
\end{displaymath}
where here and in the following we use the notation $\hateq$ to denote
equalities which hold on the boundary $\partial B_R$ only.

We analyze these boundary conditions using the harmonic decomposition
of the previous section. As noted before, it is sufficient to consider
the even parity sector. In view of
Eqs. (\ref{Eq:Psi0Harmonic}-\ref{Eq:Psi4Harmonic}), we define the
radial Weyl scalars $\psi_0 \equiv 2(e_2 - g_2)$, $\psi_1 \equiv e_1 -
g_1$, $\psi_2 \equiv e_0$, $\psi_3 \equiv e_1 + g_1$, and $\psi_4
\equiv 2(e_2 + g_2)$ which are functions of $t$ and $r$ only. Using
the relations
(\ref{Eq:e0},\ref{Eq:e1solved},\ref{Eq:e2solved},\ref{Eq:g2solved}),
the master equation (\ref{Eq:MasterEvol}), and
Eq. (\ref{Eq:P0P1Q2Sol}), we find
\begin{eqnarray}
\psi_0 &=& \frac{b_-^2\phi}{(\ell-1)\ell(\ell+1)(\ell+2) r^4}
              + \frac{8r\dot{\pi} + 5r\pi' + 6\pi}{4(\ell-1)(\ell+2)}\; ,
\label{Eq:Psi0Even}\\
\psi_1 &=& \frac{b_-\phi}{\ell(\ell+1)r^3} + \frac{\pi}{2}\; ,
\label{Eq:Psi1Even}\\
\psi_2 &=& \frac{\phi}{r^2}\; ,
\label{Eq:Psi2Even}\\
\psi_3 &=& \frac{b_+\phi}{\ell(\ell+1)r^3} - \frac{\pi}{2}\; ,
\label{Eq:Psi3Even}\\
\psi_4 &=& \frac{b_+^2\phi}{(\ell-1)\ell(\ell+1)(\ell+2) r^4}
              + \frac{-8r\dot{\pi} + 5r\pi' + 6\pi}{4(\ell-1)(\ell+2)}\; ,
\label{Eq:Psi4Even}
\end{eqnarray}
where we have introduced the operators $b_\pm = r^2(\partial_t \mp
\partial_r)$ and again assumed that $h \equiv 0$ for simplicity. In
what follows, the expressions
\begin{eqnarray}
(b_-)^m\phi_{\nwarrow,\ell}(t,r) 
 &=& \frac{(\ell+m)!}{(\ell-m)!}\, r^m \sum\limits_{j=0}^{\ell+m} (-1)^{j+m}
     \frac{(2\ell-j)!}{(\ell+m-j)!\, j!} (2r)^{j-\ell} V^{(j)}_\ell(r+t),
\qquad m=0,1,2,...\ell,
\label{Eq:b-jSolFormIn1}\\
(b_-)^m\phi_{\nwarrow,\ell}(t,r) 
 &=& (-1)^\ell r^m \sum\limits_{j=0}^{m-\ell-1} \frac{(\ell+m)!}{(2\ell+1-j)!}
     \frac{(m-\ell-1)!}{(m-\ell-j-1)!\, j!} (2r)^{\ell+1+j} V^{(j)}_\ell(r+t),
~ m=\ell+1,\ell+2,...,
\label{Eq:b-jSolFormIn2}\\
(b_-)^m\phi_{\nearrow,\ell}(t,r) 
 &=& r^m\sum\limits_{j=0}^{\ell-m} (-1)^{j+m}
     \frac{(2\ell-j)!}{(\ell-m-j)!\, j!} (2r)^{j-\ell} U^{(j)}_\ell(r-t),
\qquad m=0,1,2...\ell,
\label{Eq:b-jSolFormOut1}\\
(b_-)^m \phi_{\nearrow,\ell}(t,r) &=& 0,
\qquad m=\ell+1,\ell+2,...
\label{Eq:b-jSolFormOut2}
\end{eqnarray}
will be useful. They can be derived from the explicit expressions
(\ref{Eq:ExplicitSolFormIn},\ref{Eq:ExplicitSolFormOut}) by induction
in $m$. Corresponding expressions for $(b_+)^m\phi$ can be obtained by
flipping the sign of $t$ and interchanging $\phi_{\nwarrow,\ell}$ and
$\phi_{\nearrow,\ell}$. We see from these expressions that if $\pi =
0$ and $\phi = \phi_{\nearrow,\ell}$, then along the outgoing null
rays $r-t = const.$, we have $b_-\phi = O(r^0)$, $b_-^2\phi = O(r^0)$
and $b_+\phi = O(r^2)$, $b_+^2\phi = O(r^4)$. Therefore, the radial
Weyl scalars obey
\begin{equation}
\psi_s = O(r^{s-4}), \qquad r-t = const.
\end{equation}
for $s=0,1,2,3,4$. This is consistent with the peeling theorem
\cite{rP65}.

Next, we construct exact solutions of the IBVP described by the
evolution equations
(\ref{Eq:MasterConstraintEvol},\ref{Eq:MasterEvol}), initial data for
$\pi$ and $\phi$ on the interval $(0,R)$, and the boundary conditions
$\partial_t\Psi_0 \hateq \partial_t\Psi_1 \hateq 0$ imposed on a
sphere of radius $R > 0$. These solutions have the property of
satisfying the constraints initially ({\em ie.} $\pi(0,r) = 0$ and
$\dot{\pi}(0,r) = 0$ for all $0 < r < R$), but violating the
constraints at later times. For the sake of avoiding unnecessary
complications, we restrict ourselves to the case $\ell=2$, although
one should be able to construct similar solutions for higher $\ell$.

To construct these solutions, we start with a smooth function $F:
(0,\infty) \to \Real$ which is zero on the interval $(0,R)$ but
non-zero for $r > R$, and set
\begin{displaymath}
\pi(t,r) = a_2^\dagger a_1^\dagger F^{(5)}(r + ct),
\end{displaymath}
where $F^{(5)}$ denotes the fifth derivative of $F$. By construction,
this solves the constraint master equation
(\ref{Eq:MasterConstraintEvol}), and $\pi(0,r) = 0$, $\dot{\pi}(0,r) =
0$ for all $0 < r < R$. Choosing $h\equiv 0$ guarantees that the
constraint variables $P_0$, $P_1$ and $Q_2$ have trivial initial data
as well. Next, using the construction procedure outlined in the
previous section, we obtain the general solution of the
inhomogeneous wave equation (\ref{Eq:MasterEvol}). The result is
\begin{equation}
\phi(t,r) = a_2^\dagger a_1^\dagger
\left[ \phi_{\nearrow}(r-t) + \phi_{\nwarrow}(r+t) 
 + 24 F^{(3)}(r + ct) - 6r F^{(4)}(r + ct) \right],
\label{Eq:phiGeneral}
\end{equation}
where $\phi_{\nearrow}$ and $\phi_{\nwarrow}$ are (up to this point)
arbitrary smooth functions. In order to determine these functions, we
insert the general solution (\ref{Eq:phiGeneral}) into the boundary
conditions $\partial_t\Psi_0 \hateq \partial_t\Psi_1 \hateq 0$. Using
the expressions (\ref{Eq:Psi0Even},\ref{Eq:Psi1Even}), we obtain
\begin{eqnarray}
4r^4\psi_0(t,r) &=& \phi_{\nearrow}(r-t) + \phi_{\nwarrow}(r+t)
 - 2r\phi^{(1)}_{\nwarrow}(r+t) + 2r^2\phi^{(2)}_{\nwarrow}(r+t)
 - \frac{4}{3}\, r^3 \phi^{(3)}_{\nwarrow}(r+t) 
 + \frac{2}{3}\, r^4 \phi^{(4)}_{\nwarrow}(r+t)
\nonumber\\
 &+& 4\left[ 6 F^{(3)}(r+ct) - 9r F^{(4)}(r+ct) + 6r^2 F^{(5)}(r+ct) 
 - 2r^3 F^{(6)}(r+ct) \right],
\label{Eq:psi0Exp}\\
-r^4\psi_1(t,r) &=& \phi_{\nearrow}(r-t) - \frac{1}{2}\,r \phi_{\nearrow}(r-t)
 + \phi_{\nwarrow}(r+t) - \frac{3}{2}\, r\phi^{(1)}_{\nwarrow}(r+t)
 + r^2 \phi^{(2)}_{\nwarrow}(r+t) 
 - \frac{1}{3}\, r^3 \phi^{(3)}_{\nwarrow}(r+t)
\nonumber\\
 &+& 24 F^{(3)}(r+ct) - 30r F^{(4)}(r+ct) + \frac{33}{2}\, r^2 F^{(5)}(r+ct) 
 - 5r^3 F^{(6)}(r+ct) + r^4 F^{(7)}(r + ct).
\end{eqnarray}
The combination $B \equiv r^4\psi_1 + 4r^4\psi_0 + 2r^5\dot{\psi}_0$ gives
\begin{equation}
B(t,r) = \frac{1}{3}\, r^5\phi^{(5)}_{\nwarrow}(r+t)
 - \frac{3}{2}\, r^2 F^{(5)}(r+ct) + 3 r^3 F^{(6)}(r+ct) 
 - 3 r^4 F^{(7)}(r+ct). 
\label{Eq:BCombination}
\end{equation}
Therefore, the boundary conditions $\psi_0(t,R) = \psi_1(t,R) = 0$ for
all $t > 0$ imply $B(t,R) = 0$ for all $t > 0$. After integrating
Eq. (\ref{Eq:BCombination}) and setting five integration constants to
zero, this condition yields
\begin{displaymath}
\phi_{\nwarrow}(R+t) = \frac{144}{R^3}
 \left[ F(R+ct) - 2R F^{(1)}(R+ct) + 2R^2 F^{(2)}(R+ct) \right]
\end{displaymath}
for all $t > R$, thus determining $\phi_{\nwarrow}$ on the interval
$(R,\infty)$. On $(0,R]$ we simply set $\phi_{\nwarrow}$ to zero which
means that initially, the solution does not contain any ingoing
radiation. Plugging this and the expression (\ref{Eq:psi0Exp}) into
the boundary condition $\psi_0(t,R) = 0$ for all $t > 0$ fixes
$\phi_{\nearrow}(R-t)$ for all $t > 0$. The final result is
\begin{eqnarray}
\phi_{\nearrow}(r-t) &=& \frac{2}{R^3}
 \left[ -72F(z) + 216R F^{(1)}(z) - 324R^2 F^{(2)}(z)
   + 216R^3 F^{(3)}(z) \right. \nonumber\\
 && \left. \qquad -\, 81 R^4 F^{(4)}(z) 
  + 18R^5 F^{(5)}(z) - 2R^6F^{(6)}(z) \right]_{z = c(3R - r + t)}\; ,
\\
\phi_{\nwarrow}(r+t) &=& \frac{144}{R^3}
 \left[ F(z) - 2R F^{(1)}(z) + 2R^2 F^{(2)}(z) \right]_{z = c(R + r + t)}\; .
\end{eqnarray}
Therefore, we have constructed explicit solutions which satisfy the
boundary conditions $\partial_t\Psi_0 \hateq \partial_t\Psi_1 \hateq
0$ obtained by freezing the incoming characteristic fields of the
symmetric hyperbolic system (\ref{Eq:EvolWeylE},\ref{Eq:EvolWeylH}).
These solutions have the property that the constraints $P_a = Q_a = 0$
are satisfied initially, but violated for $t > 0$, thus providing an
explicit example which shows that freezing the incoming characteristic
fields to their initial values at the boundary is not always
compatible with constraint propagation. This fact has also been
observed in numerical simulations \cite{lKlLmSlBhP05,oSmT05}.

\subsection{Constraint-preserving boundary conditions: Freezing $\Psi_0$}
\label{SubSect:CPBCFreezingPsi0}

Here, we improve the boundary conditions considered in the previous
section. Our goal is to formulate the evolution problem in such a way
that solutions belonging to constraint-satisfying initial data
automatically satisfy the constraints everywhere on $B_R$ and at each
time $t > 0$. There are two ways to achieve this. The first approach
\cite{hFgN99} modifies the evolution equations by adding suitable
combinations of the constraint equations to them in such a way that
the resulting constraint propagation system is symmetric hyperbolic
and does not contain any normal derivatives at the
boundary. Consequently, the constraint-preserving property of the
boundary conditions is automatic. The second approach leaves the
evolution equations unchanged, but replaces the boundary condition
$\partial_t\Psi_1 \hateq 0$ with a carefully chosen boundary condition
which guarantees constraint propagation.

In the first approach, one chooses \cite{hFgN99}
\begin{eqnarray}
R_{ab} &=& +s_{(a} \varepsilon_{b)}{}^{cd} s_c Q_d\; ,
\label{Eq:RFN}\\
S_{ab} &=& -s_{(a} \varepsilon_{b)}{}^{cd} s_c P_d\; ,
\label{Eq:SFN}
\end{eqnarray}
instead of $R_{ab} = S_{ab} = 0$ in the evolution equations
(\ref{Eq:EvolWeylE},\ref{Eq:EvolWeylH}), where $P_a$ and $Q_a$ are
given by Eqs. (\ref{Eq:DivE}) and (\ref{Eq:DivH}), respectively. One
can verify that the resulting evolution system is symmetrizable
hyperbolic and that the characteristic speeds and fields with respect
to $s_a$ are unchanged, with the exception of one important
difference. The presence of the term $\pounds_s\bar{E}_a$ in
$\bar{P}_a$ (see Eq. (\ref{Eq:PbaraDef})) cancels the corresponding
term in the evolution equation for $\bar{H}_a$. Similarly, the term
$\pounds_s\bar{H}_a$ is canceled in the evolution equation for
$\bar{E}_a$. This changes the speeds of the characteristic fields
$\bar{V}^{(-)}_a$ and $\bar{V}^{(+)}_a$ from $\pm\alpha/2 - \beta^a
s_a$ to $-\beta^a s_a$. For our coordinate choice, with $\beta^a = 0$,
$\Psi_1$ is no longer an incoming field, and therefore, no longer
requires boundary data. The only remaining boundary condition is the
one involving $\Psi_0$. The definitions (\ref{Eq:RFN}) and
(\ref{Eq:SFN}) yield a different constraint propagation system than
discussed in the previous subsection. Using
Eqs. (\ref{Eq:Pbar},\ref{Eq:Pbara}), we obtain
\begin{eqnarray}
\pounds_n \bar{P} &=& 
 -\frac{1}{\alpha^2}\,\varepsilon^{ab}{\cal D}_a\left( \alpha^2\bar{Q}_b\right)
 + \frac{1}{2}\left(\bar{k} - 3k \right)\bar{P},
\label{Eq:PbarFN}\\
\pounds_n \bar{P}_a &=& 
 -\frac{1}{2\alpha^3}\,\varepsilon_a{}^b {\cal D}_b(\alpha^3\bar{Q})
 - \hat{\kappa}_{ab}\varepsilon^{bc}\bar{Q}_c
 + \left( \frac{\pounds_s\alpha}{\alpha} - \frac{\kappa}{2} \right)
   \varepsilon_a{}^b\bar{Q}_b
 - k \bar{P}_a + 2\hat{k}_a{}^b\bar{P}_b\; . 
\label{Eq:PbaraFN}
\end{eqnarray}
The corresponding equations for $\bar{Q}$ and $\bar{Q}_a$ are obtained
from this by applying the Dirac duality transformations
(\ref{Eq:DiracDuality}). For the case of linearization about Minkowski
space in the natural foliation where $\alpha = 1$, $\beta^a = 0$,
$k_{ab} = 0$, $\kappa = 2/r$, and $\hat{\kappa}_{ab} = 0$, we obtain,
using the harmonic decomposition
(\ref{Eq:PHarmDecomp},\ref{Eq:QHarmDecomp}),
\begin{eqnarray}
\dot{P}_0 &=& -\frac{\ell(\ell+1)}{r}\, Q_2\; ,
\label{Eq:P0FN}\\
\dot{P}_1 &=& \frac{1}{r} Q_2\; ,
\label{Eq:P1FN}\\
\dot{Q}_2 &=& \frac{1}{2r}\left( 2P_1 + P_0 \right).
\label{Eq:Q2FN}
\end{eqnarray}
These equations are ordinary differential equations in
time. Therefore, initial data which satisfies the constraints
automatically yield constraint-satisfying solutions.

In the second approach, one sets $R_{ab} = S_{ab} = 0$ as before, but
replaces the freezing boundary condition $\partial_t\Psi_1 \hateq 0$
with CPBC which can be constructed as follows. For $R_{ab} = S_{ab} =
0$, the constraint propagation system (\ref{Eq:Pbar},\ref{Eq:Pbara})
is a symmetric hyperbolic system whose characteristic speeds and
fields with respect to the radial field $s_a$ are
\begin{eqnarray}
& \mu = -\beta^a s_a\; , & \bar{P}, \qquad \bar{Q},\\
& \mu = -\frac{\alpha}{2} - \beta^a s_a\; , 
& \bar{W}^{(-)}_a = \bar{P}_a + \varepsilon_a{}^b \bar{Q}_b\; ,\\
& \mu = +\frac{\alpha}{2} - \beta^a s_a\; ,
& \bar{W}^{(+)}_a = \bar{P}_a - \varepsilon_a{}^b \bar{Q}_b\; .
\end{eqnarray}
If $\beta^a = 0$ (or more generally, if $\beta^a s_a \leq 0$), the
homogeneous boundary condition
\begin{displaymath}
\bar{W}_a^{(-)} \hateq 0
\end{displaymath}
guarantees that the unique solution to the constraint propagation
system with zero initial data is zero. In terms of the constraint
variables $h(r)$ and $\pi(t,r)$ introduced in
Eq. (\ref{Eq:P0P1Q2Sol}), this boundary condition yields
\begin{displaymath}
\frac{1}{c}\dot{\pi} + \pi' \hateq \frac{1}{c} h(R),
\end{displaymath}
guaranteeing that solutions of the constraint master equation
(\ref{Eq:MasterConstraintEvol}) which satisfy $h(R) = 0$, $\pi(0,r) =
0$, and $\dot{\pi}(0,r) = 0$ for $0 < r < R$ are trivial. Therefore,
there cannot exist solutions with constraint-satisfying initial data
which violate the constraints at some $t > 0$, as we encountered in the
previous subsection.

\begin{figure}[htb]
\centerline{
\includegraphics[width=8cm]{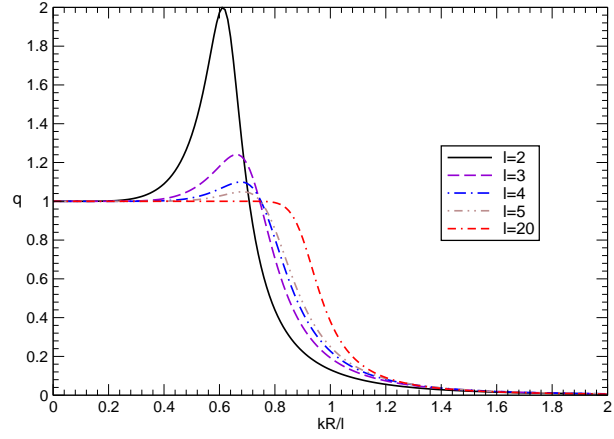}}
\vspace{0.25cm}
\caption{Reflection coefficient $q_\ell = |\gamma_\ell(kR)|$ as a
function of $k R/\ell$ ($\ell = 2,3,4,5,20 $), 
for the boundary condition $\partial_t\Psi_0 \hateq 0$.
\vspace{1cm}}
\label{Fig:Reflection_la}
\end{figure}

From now on, we assume that the constraints $P_0 = P_1 = Q_2 = 0$ are
exactly satisfied, and analyze solutions to the Bianchi equations on
$[0,T] \times B_R$ which satisfy the boundary condition
\begin{equation}
\partial_t\psi_0 \hateq 0.
\label{Eq:FrozenPsi0}
\end{equation}
As indicated above, $\psi_0$ does not vanish exactly at a finite
radius for the purely outgoing solutions $\phi_{\nearrow,\ell}$, but
falls off as $r^{-4}$ on the outgoing null rays
$r-t=const$. Therefore, imposing the boundary condition
(\ref{Eq:FrozenPsi0}) at finite radius $r=R < \infty$ yields
reflections of gravitational radiation. In other words, solutions to
the IBVP will consist of a superposition of a purely outgoing and a
purely ingoing solution. In order to quantify the amount of
reflection, we first consider monochromatic quadrupolar waves of the
form
\begin{equation}
\phi(t,r) 
 = a_2^\dagger a_1^\dagger \left( e^{ik(r-t)} + \gamma e^{-ik(r+t)} \right),
\label{Eq:InOutPhiQuad}
\end{equation}
where $k$ is a given wave number which is assumed to be different from
zero and $\gamma$ a (yet unknown) amplitude reflection
coefficient. Introducing this ansatz into the boundary condition
(\ref{Eq:FrozenPsi0}) yields
\begin{equation}
1 + \gamma\left[ 1 + 2i k R - 2 (k R)^2 - \frac{4i}{3} (k R)^3 
 + \frac{2}{3}(k R)^4 \right] e^{-2i k R} \hateq 0.
\end{equation}
As is easy to verify, the expression inside the bracket is never
zero. Therefore, we can solve this equation for $\gamma$. The
amount of reflection is given by
\begin{equation}
|\gamma_2(kR)| 
 = \left[ 1 - \frac{8}{9}(k R)^6 + \frac{4}{9}(k R)^8 \right]^{-1/2},
\label{Eq:gamma2}
\end{equation}
where the subindex $2$ refers to the fact that we are considering
quadrupolar waves. The reflection coefficient $|\gamma_2(kR)|$ versus
$kR/2$ is plotted in Figure \ref{Fig:Reflection_la}. There is a global
maximum at $k R = \sqrt{3/2}$ where $|\gamma_2(kR)| = 2$. For $k R \gg
1$, $|\gamma_2(kR)|$ decays as $(k R)^{-4}$. Therefore, the boundary
conditions are very accurate provided the size of the domain is much
larger than the characteristic wavelength of the problem. On the other
hand, if the size of the domain is comparable to the characteristic
wavelength, the reflection coefficient is of the order of unity. How
to improve this boundary condition is explained in the next subsection.

Using the expressions
(\ref{Eq:b-jSolFormIn1},\ref{Eq:b-jSolFormOut1}), the above analysis
can be repeated for arbitrary $\ell\geq 2$. The result is
\begin{equation}
|\gamma_\ell(kR)| 
 = \Big| \frac{p_{\ell,-2}(-i k R)}{p_{\ell,2}(i k R)} \Big|
\label{Eq:ReflCoeffArbitraryl}
\end{equation}
where the polynomials $p_{\ell,m}(z)$, $|m| \leq \ell$, are given by
\begin{equation}
p_{\ell,m}(z) = \sum\limits_{j=0}^{\ell+m}
  \frac{(\ell+m)!\, (2\ell-j)!}{(\ell+m-j)!\, j!}\, (2z)^j.
\label{Eq:pPolynomials}
\end{equation}
The reflection coefficients $q_\ell = |\gamma_\ell(kR)|$ versus
$kR/\ell$ for different values of $\ell$ are shown in Figure
\ref{Fig:Reflection_la}. It can be seen that $q_\ell$ is of the order
of unity for $k R/\ell < 1$ while for $k R/\ell \gg 1$, $q_\ell$
decays very rapidly. From Eq. (\ref{Eq:ReflCoeffArbitraryl}) it
follows that for large $k R$, $|\gamma_\ell(k R)|$ decays as $(k
R)^{-4}$. Although for fixed $k R$, the reflection coefficient gets
larger when $\ell$ is increased, this is not an issue for most
physically interesting scenarios, since the first few multipoles
usually dominate. In particular, if the solution is smooth, amplitudes
corresponding to different values of $\ell$ decay rapidly as
$\ell\to\infty$. Therefore, even though for high $\ell$'s the
reflection coefficient is large, it does not introduce a large overall
error since the corresponding amplitudes of the solutions should be
very small.

\begin{figure}[htb]
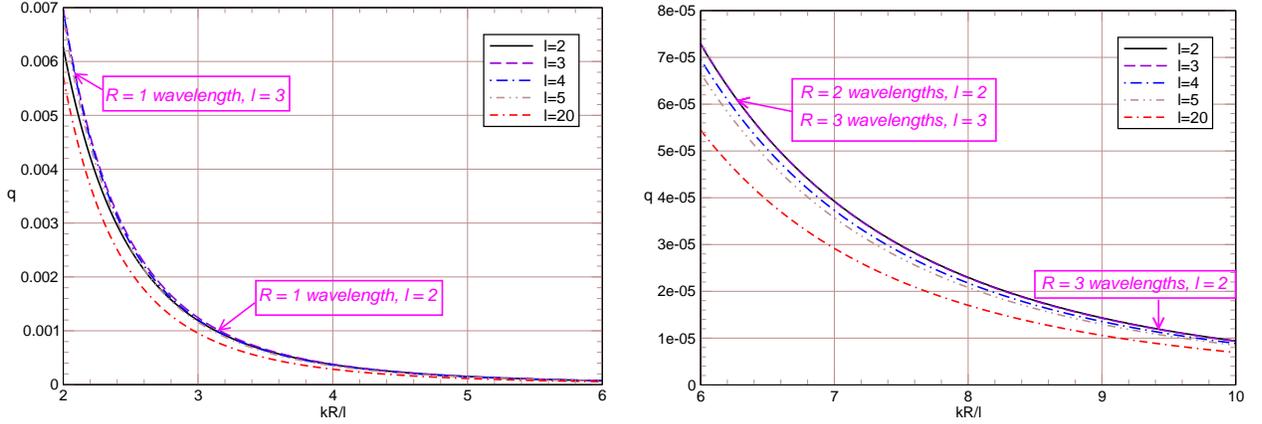

\centerline{
\includegraphics[width=8cm]{reflection_lb.eps}
\hspace{0.25cm}
\includegraphics[width=8cm]{reflection_lc.eps}}
\vspace{0.25cm}
\caption{Close-up of regions $2 \le k R/\ell \le 6 $ and
$6 \le k R/\ell \le 10 $.}
\label{Fig:Reflection_lbc}
\end{figure}

Figure \ref{Fig:Reflection_lbc} shows in more detail the amount of
reflection if the outer boundary is placed at a few multiples of the
characteristic wavelength of the problem. Clearly, this amount of
reflection is very small ($0.1\%$ or less for $R$ greater than or
equal to one wavelength and $\ell=2$, and less than $0.0065\%$ for $R$
greater than or equal to two wavelengths and $\ell=2$).

\vspace{1cm}

\vspace{1cm}

\subsection{Improved constraint-preserving boundary conditions}
\label{SubSect:CPBCImproved}

As we have seen in the previous subsection, the boundary condition
(\ref{Eq:FrozenPsi0}) is not perfectly absorbing. If the outer
boundary is a sphere of radius $R$, and for monochromatic waves with
wave number $k$, there are reflections where the reflection
coefficient is proportional to $(k R)^{-4}$ for large $k R$. Although
these reflections can be made arbitrarily small if the boundary is
pushed sufficiently far away, there is significant motivation for
improving the boundary conditions. In particular, it may not always be
possible to push the outer boundary far into the wave zone in
numerical simulations, especially for those in three space
dimensions, because of the high computational cost. Even if this can
be achieved, it may still be desirable to decrease the artificial
reflection in order to achieve better accuracy.

Our goal here is to find boundary conditions which are {\em perfectly
absorbing} at least for all multipoles $\ell=2,3,...L$, where $L$ is a
given maximum. This means that for initial data which is supported on
the interval $(0,R)$ and which corresponds to a purely outgoing
solution $\phi_{\nearrow,\ell}(t,r)$, the solution of the IBVP for $t
> 0$ is uniquely given by $\phi_{\nearrow,\ell}(t,r)$.

One way to achieve this goal is to rely on the identities (\ref{Eq:ComRel})
and the fact that $\phi_{\nearrow,\ell}$ solves the homogeneous master
equation (\ref{Eq:MasterEvol}) for each $\ell\in\Natural$. Using this
approach, we find
\begin{displaymath}
a_1 a_2 ... a_\ell \phi_{\nearrow,\ell}
 = a_1 a_2 ... a_\ell a_\ell^\dagger \phi_{\nearrow,\ell-1}
 = -a_1 a_2 ... a_{\ell-1} \partial_t^2 \phi_{\nearrow,\ell-1}
 = ... = (-1)^\ell (\partial_t)^{2\ell}\phi_{\nearrow,0}(r-t).
\end{displaymath}
This expression vanishes identically if we apply the operator $b_- =
r^2(\partial_t + \partial_r)$ to both sides. Therefore, a candidate
for our perfectly absorbing boundary condition on the field $\phi =
r^2 e_0$ is
\begin{equation}
b_- a_1 a_2 ... a_\ell\phi \hateq 0.
\end{equation}
However, a problem with this condition is that it is only {\em
quasi-local} in the sense that it is different for each
$\ell$. Therefore, a numerical implementation of the IBVP requires
performing a harmonic decomposition of the electric and magnetic
fields $E_{ab}$ and $H_{ab}$ near the outer boundary so that $\phi$
can be computed and the boundary condition applied.

An alternative approach is based on the observation that for all
$\ell\geq 2$, the outgoing solutions $\phi_{\nearrow,\ell}$ satisfy
\begin{equation}
(b_-)^{\ell+1} \phi_{\nearrow,\ell}(t,r) = 0,
\end{equation}
which follows directly from Eq. (\ref{Eq:b-jSolFormOut2}). We therefore
impose the boundary condition $(b_-)^{L+1}\phi \hateq 0$. It turns out
that this boundary condition agrees precisely with the hierarchy of
conditions given in \cite{aBeT80} for the flat wave equation in three
space dimensions. There, it was also shown that the boundary
conditions yield a well posed IBVP for the wave equation and that the
error with respect to the solution on the unbounded domain (measured
in an appropriate norm) decays as $R^{-(L+3/2)}$ as the radius $R$ of
the outer boundary goes to infinity. In order to allow for a static
contribution to $\phi$, we impose the boundary condition
\begin{equation}
(b_-)^{L+1}\partial_t\phi \hateq 0.
\label{Eq:LPerfectBC}
\end{equation}
In the appendix, we prove by deriving a suitable estimate that the
resulting IBVP is stable, and that the initial data uniquely determine
the solutions. As a consequence, the boundary condition
(\ref{Eq:LPerfectBC}) is {\em perfectly absorbing} for all multipolar
waves with $\ell\leq L$. In view of Eq. (\ref{Eq:Psi0Even}) this
boundary condition is equivalent to the condition
\begin{equation}
(b_-)^{L-1} (r^4\partial_t\psi_0) \hateq 0
\label{Eq:LPerfectBCForPsi0}
\end{equation}
on the radial Weyl scalar $\psi_0$, provided that $L \geq 1$ and that
the constraints are satisfied. Therefore, for $L \geq 1$, the boundary
conditions (\ref{Eq:LPerfectBC}) can be reformulated as boundary
conditions on the incoming characteristic field $\Psi_0$ and its
derivatives. This sheds some light onto the meaning of the freezing
$\Psi_0$ boundary condition: it is the first member of a sequence of
boundary conditions with increasing order of accuracy. By
construction, the boundary condition (\ref{Eq:LPerfectBCForPsi0}) is
exactly satisfied for all outgoing linear gravitational waves with
$\ell \leq L$. The uniqueness result in the appendix also implies that
it sets to zero any incoming gravitational radiation. Furthermore, the
boundary conditions (\ref{Eq:LPerfectBCForPsi0}) is {\em local} in the
sense that it does not depend on $\ell$. Thus, a numerical
implementation does not require a multipolar decomposition.

Finally, we compute the amount of artificial reflections for solutions
with $\ell > L$. In order to do so, we generalize the ansatz
Eq. (\ref{Eq:InOutPhiQuad}) to arbitrary $\ell$:
\begin{displaymath}
\phi(t,r) = a_\ell^\dagger ... a_1^\dagger 
   \left( e^{ik(r-t)} + \gamma e^{-ik(r+t)} \right).
\end{displaymath}
Inserting this into Eq. (\ref{Eq:LPerfectBCForPsi0}), using Eqs.
(\ref{Eq:Psi0Even},\ref{Eq:b-jSolFormIn1},\ref{Eq:b-jSolFormOut1}), and
assuming that the constraint variable $\pi$ is zero, we obtain
\begin{equation}
|\gamma_{L,\ell}(kR)| 
 = \Big| \frac{p_{\ell,-(L+1)}(-i k R)}{p_{\ell,(L+1)}(i k R)} \Big|,
\qquad \ell > L,
\end{equation}
where the polynomials $p_{\ell,m}(z)$ are given in
Eq. (\ref{Eq:pPolynomials}). In particular, $|\gamma_{L,\ell}(kR)|$
falls off as $(k R)^{-2(L+1)}$ for large $k R$.

\section{Effects due to the backscattering}
\label{Sect:Backscattering}

In this section, we want to analyze how the results obtained in
Sects. \ref{SubSect:CPBCFreezingPsi0} and \ref{SubSect:CPBCImproved}
are modified if instead of considering linear wave propagation on a
flat spacetime, the background is curved. In particular, we are
interested in a physical situation involving a localized region of
space where strong gravitational interactions take place, and where
outside this region, the gravitational field decays rapidly to flat
space. Therefore, far from the strong field region, we can expect
spacetime to be accurately described by a perturbed Schwarzschild
metric of mass $M$ representing the total mass of the system. We place
a spherical boundary of radius $r=R$, where $r$ is the area radius of
the Schwarzschild background, and assume that $2M/R \ll 1$. In the
following, we generalize the constructed in- and outgoing wave
solutions to include first order corrections in $2M/R$, and then
compute the first order correction terms to the reflection
coefficients found in the previous section. For simplicity, and since
we are only interested in the qualitative behavior of the correction
terms, we restrict ourselves to perturbations with odd parity. The
effects of second order corrections in $2M/R$, corrections due to
$J/R^2$, where $J$ is the total angular momentum of the system, and
corrections emanating from nonlinearities (see
Ref. \cite{nBrGlLbSjWrI98} for an estimate on the errors introduced by
neglecting the nonlinearities of the theory) will be considered
elsewhere.

\subsection{Odd-parity linear fluctuations and derivation of a master equation for $\im\Psi_2$}

As shown in Ref. \cite{rP72b}, linear odd-parity metric perturbations
about a Schwarzschild black hole can be described by a master equation
for $\im\delta\Psi_2$, where $\delta\Psi_2$ denotes the linearization
of $\Psi_2$. Since $\im\Psi_2$ is a scalar field that vanishes on a
spherically symmetric background with an adapted Newman-Penrose null
tetrad, its perturbation is invariant with respect to infinitesimal
coordinate transformations. Additionally, one can also show
\cite{rP72b} that $\im\delta\Psi_2$ is invariant with respect to
infinitesimal rotations of the null tetrad, and is therefore
well-suited for describing odd-parity gravitational perturbations. It
turns out that the master equation for this quantity is the
Regge-Wheeler equation \cite{tRjW57}. In this subsection, we briefly
review the derivation of the Regge-Wheeler equation for
$\im\Psi_2$. For simplicity, we assume that the background is written
in standard Schwarzschild coordinates $(t,r,\vartheta,\varphi)$ for
which
\begin{eqnarray}
&& n^a\partial_a = \frac{1}{\alpha}\partial_t\; , \qquad
   s^a\partial_a = \alpha\partial_r\; , \qquad
   \gamma_{ab} dx^a dx^b 
     = r^2\left( d\vartheta^2 + \sin^2\vartheta\, d\varphi^2 \right),
\nonumber\\
&& k_{ab} = 0, \qquad
   \hat{\kappa}_{ab} = 0, \qquad
   \kappa = \frac{2\alpha}{r}\; ,
\nonumber
\end{eqnarray}
where $\alpha = \sqrt{1 - 2M/r}$. The corresponding electric and
magnetic parts of the Weyl tensor are
\begin{displaymath}
\Ez_{ab} = \frac{M}{r^3}( \gamma_{ab} - 2s_a s_b),
\qquad
\Hz_{ab} = 0,
\end{displaymath}
where the circles on the top of $\Ez_{ab}$ and $\Hz_{ab}$ indicate
that they are background quantities. Linearizing the evolution and
constraint equations
(\ref{Eq:EvolWeylE},\ref{Eq:EvolWeylH},\ref{Eq:DivE},\ref{Eq:DivH})
about this background yields the system
\begin{eqnarray}
\pounds_n\tilde{E}_{ab} + \varepsilon_{cd(a} (D^c + 2a^c)\tilde{H}^d{}_{b)} 
 &=& R_{ab}\; ,
\label{Eq:EvolWeylPertE}\\
\pounds_n\tilde{H}_{ab} - \varepsilon_{cd(a} (D^c + 2a^c)\tilde{E}^d{}_{b)}
 &=& S_{ab}\; ,
\label{Eq:EvolWeylPertH}\\
D^b\tilde{E}_{ab} &=& P_a\; ,
\label{Eq:DivPertE}\\
D^b\tilde{H}_{ab} &=& Q_a\; ,
\label{Eq:DivPertH}
\end{eqnarray}
where $n$, $\varepsilon_{abc}$, $D$, and $a^c$ refer to the {\em
background} geometry, and where $\tilde{E}_{ab}$ and $\tilde{H}_{ab}$
denote the parts of the perturbed electric and magnetic components of
the Weyl tensor which are trace-free with respect to the {\em
background} metric $h_{ab} = s_a s_b + \gamma_{ab}$. Also, indices are
raised and lowered with the background metric. The source terms
$R_{ab}$, $S_{ab}$, $P_a$, and $Q_a$ depend on the perturbations of
the shift, $\delta\beta^a$, the perturbations of the metric, $\delta
h_{ab}$, and its first spatial derivatives, and the perturbations of
the extrinsic curvature, $\delta k_{ab}$. Performing a change of
infinitesimal coordinates if necessary, we can obtain $\delta\beta^a =
0$. In this case, we find that the source terms $R_{ab}$, $S_{ab}$,
$P_a$, and $Q_a$ are given by
\begin{eqnarray}
R_{ab} &=& 5\left( \Ez_{(a}{}^c\delta k_{b)c} 
 - \frac{1}{3} h_{ab}\Ez^{cd}\delta k_{cd} \right)
 - 2\Ez_{ab} h^{cd}\delta k_{cd}\; ,\\
S_{ab} &=& -\varepsilon_{(a}{}^{cd}\left[ 
 \delta h_{b)c} a^e\Ez_{de} + C^e{}_{b)c}\Ez_{de} 
 + 2\Ez_{b)c} D_d \left( \frac{\delta\alpha}{\alpha} \right) \right]
\\
P_a &=& \left( D^b\Ez^c{}_a - \frac{1}{3} D_a\Ez^{bc} \right)\delta h_{bc} 
 + \Ez_{ab} h^{cd} C^b{}_{cd} + \frac{1}{3}\Ez_b{}^c C^b{}_{ca}\; ,\\
Q_a &=& -\varepsilon_a{}^{bc}\Ez_b{}^{d}\delta k_{cd}\; ,
\end{eqnarray}
where
\begin{displaymath}
C^c{}_{ab} = \frac{1}{2}h^{cd}\left( D_a\delta h_{bd} + D_b\delta h_{ad}
 - D_d\delta h_{ab} \right)
\end{displaymath}
are the linearized Christoffel symbols. Here, we have also used the
background equations
\begin{displaymath}
\pounds_n\delta h_{ab} = 2\delta k_{ab}\; ,\qquad
\varepsilon_{cd(a} (D^c + 2a^c)\Ez^d{}_{b)} = 0, \qquad
D^b \Ez_{ab} = 0,
\end{displaymath}
which imply
\begin{displaymath}
\varepsilon_{cda} (D^c + 2a^c)\Ez^d{}_b = \varepsilon_{abc} a_d\Ez^{cd}.
\end{displaymath}

Next, we perform a $2+1$ split of
Eqs. (\ref{Eq:EvolWeylPertE},\ref{Eq:EvolWeylPertH},\ref{Eq:DivPertE},\ref{Eq:DivPertH})
as described in Sect. \ref{SubSect:2+1}. Notice that the $2+1$ split
is with respect to the unperturbed Schwarzschild metric, for which the
assumptions made below Eq. (\ref{Eq:Psi4}) on the vector fields $s^a$
and $n^a$ hold, and not with respect to the perturbed Schwarzschild
metric. Using the odd-parity sector of the harmonic decomposition
(\ref{Eq:EHHarmDecomp}) for $\tilde{E}_{ab}$ and $\tilde{H}_{ab}$, and
a corresponding odd-parity decomposition for $\delta h_{ab}$ and
$\delta k_{ab}$, namely,
\begin{eqnarray}
\delta h_{ab} &=& 2\sigma(t,r) s_{(a}\hat{S}_{b)} 
               +  2r \nu(t,r) \hat{\nabla}_{(a}\hat{S}_{b)}\; ,
\nonumber\\
\delta k_{ab} &=& 2\pi_\sigma(t,r) s_{(a}\hat{S}_{b)} 
               +  2r \pi_\nu(t,r) \hat{\nabla}_{(a}\hat{S}_{b)}\; ,
\nonumber
\end{eqnarray}
we obtain the following equations for $\ell\geq 2$:
\begin{eqnarray}
\frac{\dot{h}_0}{\alpha} - \frac{\ell(\ell+1)}{r} f_1 
 &=& -\frac{\ell(\ell+1)}{r}\frac{M}{r^3}\sigma\; ,
\label{Eq:h0dot}\\
\frac{\dot{f}_1}{\alpha} - \frac{1}{2\alpha}( \alpha^2 h_1 )'
 - \frac{\lambda}{2r} h_2 + \frac{3}{4r} h_0 
 &=& -\frac{5M}{2r^3}\pi_\sigma\; ,
\label{Eq:f1dot}\\
\frac{\alpha}{r^2}(r^2 h_0)' - \frac{\ell(\ell+1)}{r} h_1 &=& 0,
\label{Eq:h1Elim}\\
\frac{\alpha}{r^2}(r^2 h_1)' - \frac{\lambda}{r} h_2 - \frac{1}{2r} h_0
 &=& \frac{3M}{r^3}\pi_\sigma\; ,
\label{Eq:h2Elim}\\
\frac{\alpha}{r^2}(r^2 f_1)' - \frac{\lambda}{r} f_2 
 &=& \frac{M}{r^3}\left[ \alpha r\left( \frac{\sigma}{r} \right)' 
  - \frac{\lambda}{r}\nu \right].
\label{Eq:f2Elim}
\end{eqnarray}

A master equation for $\phi \equiv r^2 h_0$ is obtained as follows.
First, we use Eqs. (\ref{Eq:h2Elim}) and (\ref{Eq:h1Elim}) in order to
eliminate $h_2$ and $h_1$ in Eq. (\ref{Eq:f1dot}). Then, we use
Eq. (\ref{Eq:h0dot}) and the definition of the extrinsic curvature,
$\dot{\sigma} = 2\alpha\pi_\sigma$, in order to eliminate $\dot{f}_1$
from the resulting equation. This yields
\begin{equation}
\left[ \frac{1}{\alpha^2}\partial_t^2 - \partial_r(\alpha^2\partial_r) 
 + \frac{\ell(\ell+1)}{r^2} \right]\phi(t,r) 
 = -6M\frac{\ell(\ell+1)}{r^2}\pi_\sigma\; .
\end{equation}
To get an equation for $\phi$ alone, we need a relation between
$\pi_\sigma$ and $\phi$. This is obtained by linearizing the equation
\begin{displaymath}
H_{ab} = -\varepsilon_{cd(a} D^c k^d{}_{b)},
\end{displaymath}
which expresses the magnetic part of the Weyl tensor in terms of the
curl of the extrinsic curvature. This yields
\begin{equation}
\phi = -\ell(\ell+1)r\pi_\sigma\; ,
\label{Eq:RelPhiPi}
\end{equation}
and leads to the Regge-Wheeler equation \cite{tRjW57}
\begin{equation}
\left[ \frac{1}{\alpha^2}\partial_t^2 - \partial_r(\alpha^2\partial_r) 
 + \left( \frac{\ell(\ell+1)}{r^2} - \frac{6M}{r^3} \right) \right]\phi(t,r) 
 = 0
\label{Eq:RW}
\end{equation}
for $\phi$. As in the flat spacetime case, the linearized
Newman-Penrose scalar $\delta \Psi_0$ is entirely determined by
$\phi$. To see this, we first linearize
Eq. (\ref{Eq:Psi0})\footnote{Notice that only the property that $s^a$
is a unit vector field which is everywhere orthogonal to $n^a$ was
used in the derivation of Eq. (\ref{Eq:Psi0}). Therefore, we may
assume that $s^a$ exists also for the perturbed spacetime.} and obtain
\begin{displaymath}
\delta\Psi_0 = \frac{\psi_0}{r}\, \hat{m}^A\hat{m}^B\hat{\nabla}_A\hat{S}_B\; ,
\qquad
\psi_0 \equiv 2\left( h_2 + f_2 - \frac{M}{r^3}\nu \right).
\end{displaymath}
Using Eqs. (\ref{Eq:h0dot}-\ref{Eq:f2Elim}) and
Eq. (\ref{Eq:RelPhiPi}), we re-express the above expression in terms
of $\phi$ alone, giving
\begin{equation}
\psi_0 = \frac{\alpha^2 b_-^2\phi}{(\ell-1)\ell(\ell+1)(\ell+2) r^4}\; ,
\label{Eq:Psi0Schw} 
\end{equation}
where $b_- \equiv r^2(\alpha^{-2}\partial_t + \partial_r)$. This
generalizes Eq. (\ref{Eq:Psi0Even}) to a Schwarzschild background.

\subsection{Construction of in- and outgoing wave solutions to
 first order in $M/R$}

Next, we generalize the in- and outgoing wave solutions constructed in
Sect. \ref{SubSect:ExactSolutions} to the case $M \neq 0$. This means
that we have to solve the new master equation (\ref{Eq:RW}). Since we
are only interested in cases where $M/r \ll 1$, it is reasonable to
expand the equation in factors of $M/r$ and to consider the first
order corrections in $M/r$ only. This expansion might depend on the
chosen coordinates. For the following, it is convenient to introduce
the tortoise coordinate $r_*$ which is defined by
\begin{displaymath}
r_*(r) \equiv \int\limits_{4M}^r \frac{ds}{1 - \frac{2M}{s}}
 = r - 4M + 2M\log\left( \frac{r}{2M} - 1 \right).
\end{displaymath}
Using this, we can rewrite the Regge-Wheeler equation as
\begin{equation}
\left[ \partial_t^2 - \partial_{r_*}^2 
 + \left( 1 - \frac{2M}{r} \right)
   \left( \frac{\ell(\ell+1)}{r^2} - \frac{6M}{r^3} \right) \right]\phi(t,r) 
 = 0.
\label{Eq:RWbis}
\end{equation}
Therefore, if $r$ is very large compared to $\ell$ and $M$, in- and
outgoing solutions are, approximately, given by $\phi_{\nwarrow}(t,r)
\approx V(r_* + t)$ and $\phi_{\nearrow}(t,r) \approx U(r_* - t)$,
respectively. Notice that $r_*$ is not analytic in $2M/r$ at $2M/r =
0$, so it is not clear if, for example, $\phi_{\nearrow}(t,r) \approx
U(r - t)$ ($r_*$ replaced by $r$) is a good approximation for the
behavior of outgoing solutions in the asymptotic regime. For this
reason, it seems more appropriate to use the coordinates $(t,r_*)$ to
describe the asymptotic behavior of the solutions. On the other hand,
the potential term appearing in Eq. (\ref{Eq:RWbis}) is not analytic
in $2M/r_*$ at $2M/r_* = 0$, so we cannot expand it in terms of powers
of $2M/r_*$ near $2M/r_* = 0$. In order to circumvent this problem,
we introduce the new coordinates
\begin{displaymath}
\tau = t + r - r_*\; , \qquad
\rho = r,
\end{displaymath}
in which the Regge-Wheeler equation can be written as
\begin{equation}
\left[ \partial_{\tau}^2 - \partial_{\rho}^2 
 + \frac{\ell(\ell+1)}{\rho^2} \right]\phi(\tau,\rho) 
 = -\frac{2M}{\rho} B\phi(\tau,\rho),
\label{Eq:RWBisBis}
\end{equation}
where the operator $B$ is defined by
\begin{displaymath}
B = \left( \partial_{\tau} + \partial_{\rho} - \frac{2}{\rho} \right)
    \left( \partial_{\tau} + \partial_{\rho} + \frac{1}{\rho} \right).
\end{displaymath}
If we neglect the right-hand side, this equation reduces to the flat
space master equation which has the outgoing solutions
$\phi_{\nearrow,\ell}(\tau,\rho)$ constructed in
Sect. \ref{SubSect:ExactSolutions}. These outgoing solutions have the
correct asymptotic behavior since $\phi_{\nearrow,\ell}(\tau,\rho)
\approx U^{(\ell)}_\ell(\rho - \tau) = U^{(\ell)}_\ell(r_* - t)$. We
also see that for these solutions, $B \phi_{\nearrow,\ell}(\tau,\rho)$
decays as $\rho^{-2} = r^{-2}$, so the right-hand side of
(\ref{Eq:RWBisBis}) is small. Therefore, given $R > 2M$, we expect
that we can write the solution in terms of an expansion in powers of
$2M/R$ as
\begin{equation}
\phi(\tau,\rho) = a_\ell(\rho)^\dagger a_{\ell-1}(\rho)^\dagger... 
                  a_1(\rho)^\dagger U(\rho-\tau)
 + \sum\limits_{k=1}^\infty \left( \frac{2M}{R} \right)^k g_k(\tau,\rho),
\label{Eq:2MOverRExpansion}
\end{equation}
for all $\rho$ in a neighborhood of $R$, where here and in the
following, $a_\ell(\rho)^\dagger = -\partial_\rho + \ell/\rho$. In
Ref. \cite{jBwP73}, a similar expansion was used to obtain solutions
of the Teukolsky equation \cite{Teukolsky72} on a Schwarzschild
background, and was shown to converge absolutely. Plugging the
expansion (\ref{Eq:2MOverRExpansion}) into Eq. (\ref{Eq:RWBisBis})
yields the following hierarchy of partial differential equations:
\begin{equation}
\left[ \partial_{\tau}^2 - \partial_{\rho}^2 
 + \frac{\ell(\ell+1)}{\rho^2} \right] g_k(\tau,\rho) 
 = -\frac{R}{\rho} B g_{k-1}(\tau,\rho), \qquad
k = 1,2,3,...
\label{Eq:gk}
\end{equation}
where $g_0(\tau,\rho) \equiv a_\ell(\rho)^\dagger
... a_1(\rho)^\dagger U(\rho - \tau)$. In
Sect. \ref{SubSect:ExactSolutions}, we learned how to solve such
equations using integral representations of the solution operator of
the flat wave equation.

In the following, we give the explicit solution for the first order
correction ($k=1$) of quadrupolar waves ($\ell=2$). The solution can be
written as
\begin{equation}
g_1(\tau,\rho) = \frac{3R}{4\rho^2} U^{(1)}(\rho-\tau)
 + \frac{R}{4}\int\limits_{\rho-\tau}^\infty K_2(\tau,\rho,x) U(x) dx,
\label{Eq:FirstOrderCorrection}
\end{equation}
where the integral kernel $K_2$ is given by
\begin{displaymath}
K_2(\tau,\rho,x) \equiv a_2(\rho)^\dagger a_1(\rho)^\dagger 
 \frac{4}{(\tau+\rho+x)^2}
 = \frac{3}{2\rho^4}\left[ w^{-4} + 2w^{-3} + 2w^{-2} 
 \right]_{w = \frac{\tau+\rho+x}{2\rho}}\; ,\qquad x > \rho - \tau.
\end{displaymath}
and satisfies
\begin{eqnarray}
&& \left[ \partial_{\tau}^2 - \partial_{\rho}^2 
  + \frac{6}{\rho^2} \right] K_2(\tau,\rho,x) = 0, 
\nonumber\\
&& K_2(\tau,\rho,\rho-\tau) = \frac{15}{2\rho^4}\; ,
\label{Eq:K2Properties}\\
&& (\partial_\tau + \partial_\rho) K_2(\tau,\rho,\rho-\tau) 
 = -\frac{30}{\rho^5}\; .
\nonumber
\end{eqnarray}
Notice that for $\tau > 0$ and $\rho > 0$, the function $x\mapsto
K_2(\tau,\rho,x)$ is bounded from above by the function
\begin{displaymath}
x \mapsto M_1(x) \equiv \frac{30}{\rho^2} \frac{1}{(\rho + x)^2}
\end{displaymath}
on the open interval $x > \rho - \tau$. Therefore, if $U$ is
continuous, supported on the interval $(0,\infty)$, and bounded, then
the integral in Eq. (\ref{Eq:FirstOrderCorrection}) exists for all
$\tau > 0$ and all $\rho > 0$. Using the properties
(\ref{Eq:K2Properties}), it is not difficult to verify that $g_1$
indeed solves Eq. (\ref{Eq:gk}) for $\ell=2$ and $k=1$. Notice that if
$U$ is supported in $[r_1,r_2]$, where $0 < r_1 < r_2$, the zeroth
order solution $g_0(\tau,\rho) = U^{(2)}(\rho-\tau) -
3U^{(1)}(\rho-\tau)/\rho + 3U(\rho-\tau)/\rho^2$ is supported in $[r_1
+ \tau, r_2 + \tau]$ for each $\tau > 0$. In particular, for each
fixed $\rho_1 > 0$, $g_0(\tau,\rho_1)$ vanishes for $\tau$ large
enough. This is a manifestation of Huygens' principle which holds for
the flat wave equation in odd space dimensions. The first order
correction term $g_1$ vanishes for $\rho > r_2 + \tau$, but not
necessarily for $\rho < r_1 + \tau$. This is the effect of the
backscattering. Nevertheless, for each fixed $\rho_1 > 0$,
$g_1(\tau,\rho_1)$ converges to zero as $\tau\to\infty$. This can be
shown by using Lebesgue's dominated convergence theorem\footnote{See,
for instance, chapter 4.4 in \cite{Royden}.} and noticing that the
function $x\mapsto K_2(\tau,\rho_1,x) U(x)$, which is bounded by the
integrable function $x \mapsto M_1(x) |U(x)|$ on the interval $x >
\rho_1 - \tau$, converges pointwise to zero as $\tau \to \infty$.

Summarizing, we have obtained outgoing, approximate solutions of the
Regge-Wheeler equation for $\ell=2$:
\begin{eqnarray}
\phi_{\nearrow}(t,r) &=& 
 U^{(2)}(r_*-t) - \frac{3}{r} U^{(1)}(r_*-t) + \frac{3}{r^2} U(r_*-t)
\nonumber\\
 &+& \frac{2M}{R}\left[ \frac{3R}{4 r^2} U^{(1)}(r_* - t)
  + \frac{R}{4}\int\limits_{r_* - t}^\infty K_2(t+r-r_*,r,x) U(x) dx \right]
  + O\left( \frac{2M}{R} \right)^2.
\nonumber
\end{eqnarray}
Since the Regge-Wheeler equation is time-symmetric, corresponding
ingoing solutions are obtained from this by merely flipping the sign
of $t$:
\begin{eqnarray}
\phi_{\nwarrow}(t,r) &=&
 V^{(2)}(r_*+t) - \frac{3}{r} V^{(1)}(r_*+t) + \frac{3}{r^2} V(r_*+t)
\nonumber\\
 &+& \frac{2M}{R}\left[ \frac{3R}{4 r^2} V^{(1)}(r_* + t)
  + \frac{R}{4}\int\limits_{r_* + t}^\infty K_2(-t+r-r_*,r,x) V(x) dx 
 \right] + O\left( \frac{2M}{R} \right)^2.
\nonumber
\end{eqnarray}
Using Eq. (\ref{Eq:Psi0Schw}) and the fact that $b_- =
\alpha^{-2}r^2(\partial_t + \partial_{r_*})$, we compute the
corresponding expressions for $\psi_0$:
\begin{eqnarray}
\psi_{0\nearrow}(t,r) &=& \frac{1}{4\alpha^2 r^4}\left( U(r_* - t)
 + \frac{2M}{r}\left[ -2U(r_* - t) + \frac{r}{4}U^{(1)}(r_* - t)
 + \frac{1}{2}\int\limits_0^\infty k(1+y) U(r_* - t + 2 r y) dy \right] \right)
\nonumber\\
 &+& O\left( \frac{2M}{R} \right)^2,
\label{Eq:Psi0SchwOut}\\
\psi_{0\nwarrow}(t,r) &=& \frac{1}{4\alpha^2 r^4}\left( V(r_* + t)
 - 2r V^{(1)}(r_* + t) + 2r^2 V^{(2)}(r_* + t) 
 - \frac{4}{3} r^3 V^{(3)}(r_* + t) + \frac{2}{3} r^4 V^{(4)}(r_* + t) \right.
\nonumber\\
 &+& \left. \frac{2M}{r}\left[ \frac{1}{2}r^2 V^{(2)}(r_* + t) 
 - \frac{1}{2}r^3 V^{(3)}(r_* + t)
 + \frac{1}{2}\int\limits_0^\infty \frac{V(r_* + t + 2 r y) dy}{(1+y)^2} 
\right]  \right) + O\left( \frac{2M}{R} \right)^2,
\label{Eq:Psi0SchwIn}
\end{eqnarray}
where $k(w) \equiv 5w^{-6} + 4w^{-5} + 3w^{-4} + 2w^{-3} + w^{-2}$.
Taking into account the fact that $\alpha^{-2} = 1 + 2M/r + O(2M/r)^2$,
and replacing $U(x)$ by $G(x) = U(-x) + M U^{(1)}(-x)/2$, the result
for $\alpha^{-2}\Psi_{0,\nearrow}$\footnote{In Ref. \cite{jBwP73} a
different normalization of the null vectors are used, which explains
the factor $\alpha^{-2}$.} agrees with Eq. (4.18) of
Ref. \cite{jBwP73}.

\subsection{Reflection coefficient for the boundary condition $\partial_t\Psi_0 \hateq 0$}

In order to quantify the amount of artificial reflections caused by a
spherical artificial outer boundary at $R \gg 2M$, we consider as
before monochromatic waves of the form
\begin{equation}
U(r_* - t) = e^{ik(r_*-t)}, \qquad
V(r_* + t) = \gamma e^{-ik(r_*+t)},
\label{Eq:Monochromatic}
\end{equation}
where $\gamma$ is the amplitude reflection coefficient. Introducing
these expressions into Eqs. (\ref{Eq:Psi0SchwOut},\ref{Eq:Psi0SchwIn})
and setting $\partial_t\psi_{0,\nearrow}(t,R) +
\partial_t\psi_{0,\nwarrow}(t,R) = 0$, we obtain the result
\begin{eqnarray}
\Big| \gamma_2\left( kR,\frac{2M}{R} \right) \Big| 
 &=& |\gamma_2(kR)|\left[ 1 + \frac{2M}{R} E(kR) 
 + O\left( \frac{2M}{R} \right)^2 \right],
\end{eqnarray}
where $|\gamma_2(kR)|$ is the reflection coefficient given in
(\ref{Eq:gamma2}) (which is valid for $M=0$), and the function $E(z)$
is given by
\begin{equation}
E(z) = -\frac{1}{9}\, z^6 |\gamma_2(z)|^2\left[
 8 z^2 - 13 - \left( 2 z^2 - 4 \right)
 \int\limits_0^\infty k(1+y)\cos(2z y) dy \right].
\end{equation}
In deriving this result, we have used the integrals
\begin{displaymath}
C_n(z) = \int\limits_0^\infty \frac{\cos(2zy)}{(1+y)^n}\, dy, \qquad
S_n(z) = \int\limits_0^\infty \frac{\sin(2zy)}{(1+y)^n}\, dy,
\end{displaymath}
and the relations
\begin{displaymath}
C_{n+1}(z) = \frac{1}{n}\left[ 1 - 2z S_n(z) \right], \qquad
S_{n+1}(z) = \frac{2z}{n} C_n(z),
\end{displaymath}
for $n \geq 2$, which imply that
\begin{displaymath}
\lim\limits_{z\to\infty} (2z)^2 C_n(z) = n, \qquad
\lim\limits_{z\to\infty} (2z) S_n(z) = 1,
\end{displaymath}
for all $n \geq 2$ and
\begin{displaymath}
 4 - z^2 + \left( \frac{2}{3}\, z^4 - 2z^2 + 1 \right) C_2(z)
   + \left( \frac{4}{3}\, z^3 - 2z \right) S_2(z) 
 = \int\limits_0^\infty k(1+y)\cos(2z y) dy.
\end{displaymath}
Since $E(z) \to -2$ as $z\to \infty$ it follows that
$|\gamma_2(kR,2M/R)|$ still decays as $(k R)^{-4}$ for large $k R$. In
fact, for $k R$ sufficiently large, the reflection coefficient is {\em
smaller} than the corresponding flat space coefficient provided that
$2M/R$ is small enough.

\vspace{1cm}
\begin{figure}[htb]
\centerline{
\includegraphics[width=9cm]{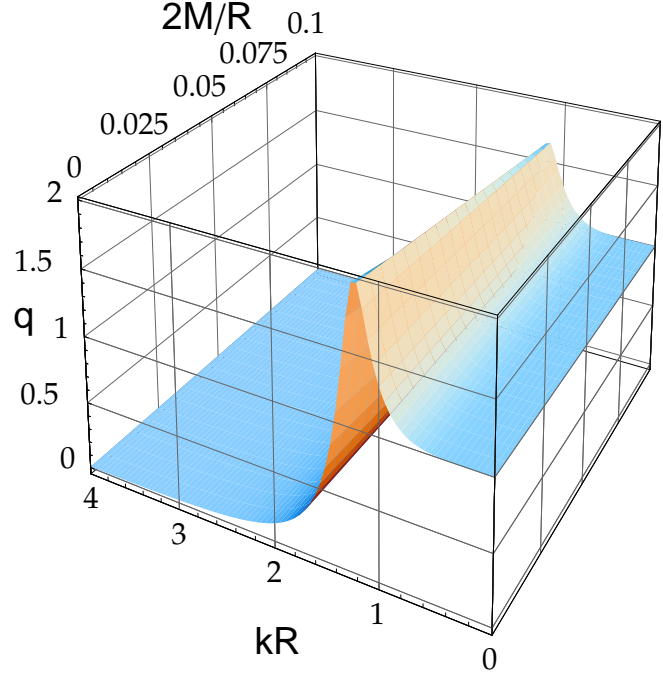}}
\vspace{0.25cm}
\caption{Reflection coefficient $q_2 = |\gamma_2(k R, 2M/R)|$
truncated to first order in $2M/R$ as a function of $k R$ and $2M/R$,
for the boundary condition $\partial_t\Psi_0 \hateq 0$.}
\label{Fig:Reflection_BHa}
\end{figure}

\vspace{1cm}
\begin{figure}[htb]
\centerline{
\includegraphics[width=9cm]{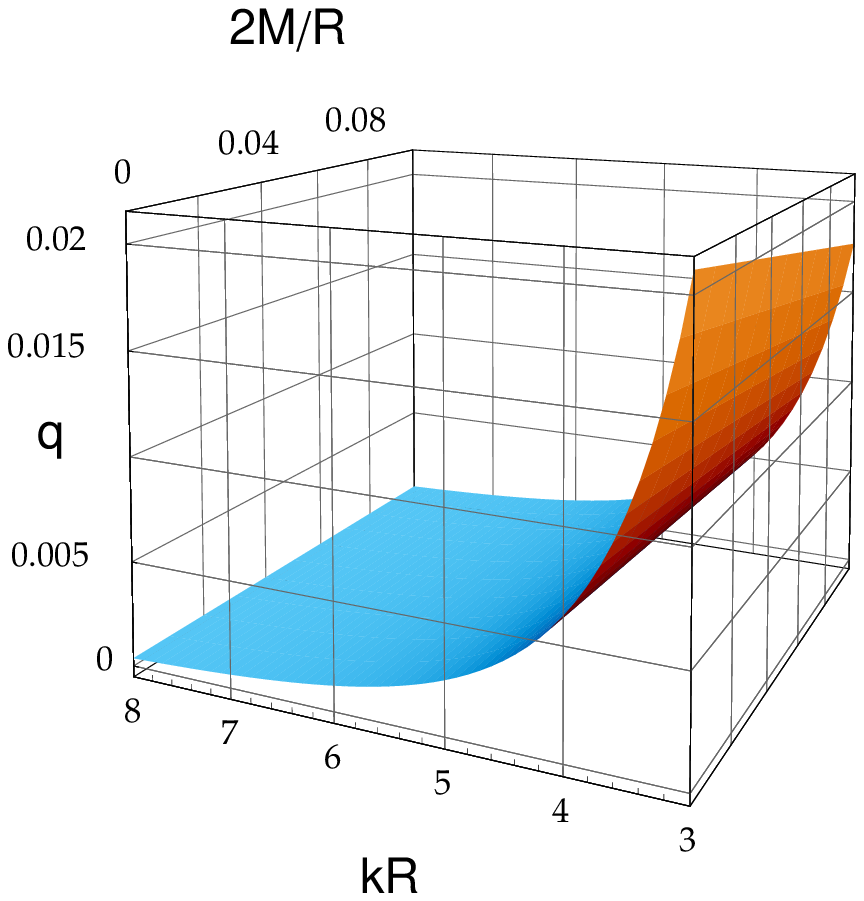}}
\vspace{0.25cm}
\caption{Reflection coefficient $q_2 = |\gamma_2(k R, 2M/R)|$
truncated to first order in $2M/R$ as a function of $k R$ and $2M/R$,
for the boundary condition $\partial_t\Psi_0 \hateq 0$. Showing
surface for $3 \le k R \le 8 $.}
\label{Fig:Reflection_BHb}
\end{figure}

\subsection{Reflection coefficient for the improved boundary condition}

\begin{figure}[htb]
\centerline{
\includegraphics[width=8cm]{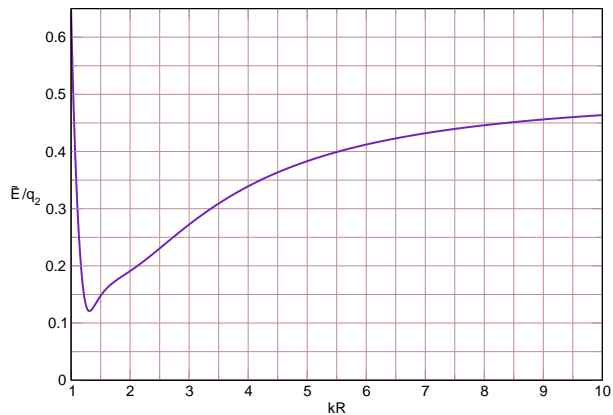}}
\vspace{0.25cm}
\caption{$\tilde{E}(kR)/|\gamma_2(kR)|$ versus $k R$.}
\label{Fig:ImprovedBC}
\end{figure}

Finally, we repeat the above analysis for the boundary condition
\begin{equation}
(\partial_t + \partial_r) (r^4\partial_t\psi_0)(t,R) = 0,
\end{equation}
which is perfectly absorbing for $M=0$ (see
Eq. (\ref{Eq:LPerfectBCForPsi0})). From
Eqs. (\ref{Eq:Psi0SchwOut},\ref{Eq:Psi0SchwIn}) we first obtain
\begin{eqnarray}
4r(\partial_t + \partial_r) (r^4\psi_{0\nearrow})(t,r)
 &=& \frac{2M}{r} \left[ U(r_* - t) + rU^{(1)}(r_* - t)
  - 15\int\limits_0^\infty \frac{U(r_* - t + 2ry) dy}{(1+y)^7} \right]
  + O\left( \frac{2M}{R} \right)^2
\nonumber\\
4r(\partial_t + \partial_r) (r^4\psi_{0\nwarrow})(t,r)
 &=& \frac{4}{3} r^5 V^{(5)}(r_* + t) + O\left( \frac{2M}{R} \right).
\nonumber
\end{eqnarray}
Using the monochromatic ansatz (\ref{Eq:Monochromatic}), we obtain
\begin{equation}
\Big| \gamma_{2,2}\left( kR,\frac{2M}{R} \right) \Big|
 = \frac{2M}{R} \tilde{E}(k R) + O\left( \frac{2M}{R} \right)^2, 
\end{equation}
where
\begin{equation}
\tilde{E}(z) = \frac{3}{4 z^5} \left[ \left( 1 - 15 C_7(z) \right)^2
 + \left( z - 15 S_7(z) \right)^2 \right]^{1/2}.
\end{equation}
For $k R \gg 1$, the reflection coefficient goes as $(2M/R)(k
R)^{-4}$. Because of the presence of the small factor $(2M/R)$, there
is a significant improvement over the boundary condition
$\partial_t\Psi_0 \hateq 0$. In Figure \ref{Fig:ImprovedBC}, we plot
the ratio $\tilde{E}(kR)/|\gamma_2(kR)|$ as a function of $k R$. This
plot, together with the asymptotic expansion
$2\tilde{E}(z)/|\gamma_2(z)| = 1 - 8z^{-2} + O(z^{-3})$, suggest that
for $k R > 1.04$, this ratio does not exceed $0.5$. Thus, we conclude
that with corrections for backscatter, our improved boundary condition
gives a reflection coefficient which is $M/R$ times smaller than the
one for the freezing $\Psi_0$ condition for $k R > 1.04$.
\vspace{1cm}

\section{Conclusions}
\label{Sect:Conclusions}

Numerical relativity groups around the world have begun to calculate
binary black hole merger waveforms
\cite{fP05,fP06,mCcLyZ06a,jBjCdCmKjvM06a,mCcLpMyZ06,jBjCdCmKjvM06b,
pDfHdPeSeSrTjTjV06,fHdSpL06,jBjCdCmKjvMmM06,mCcLyZ06b,uS06,mShPlLlKoRsT06},
with the goal of providing waveform templates for the detection and
interpretation of gravitational wave signals from instruments such as
LIGO\footnote{http://ligo.caltech.edu},
VIRGO\footnote{http://cascina.virgo.infn.it} and
LISA\footnote{http://lisa.nasa.gov}. To be useful templates, the
calculated waveforms must be as accurate as possible. In particular,
if numerical binary black hole simulations are performed on a finite
computational grid with an artificial outer boundary, it is critical
that this boundary be as seamless an interface as possible between the
physical scenario and the computational grid. Towards this end, we
have constructed a hierarchy ${\cal B}_L$, $L=1,2,3,...$ of boundary
conditions which are {\em perfectly absorbing} for linearized waves
with arbitrary angular momentum number $\ell \leq L$ on a Minkowski
background with a spherical outer boundary. For a nonlinear Cauchy
formulation of Einstein's vacuum field equations, these boundary
conditions can be formulated as follows. Let $t$ be the time-like
coordinate compatible with the foliation $\Sigma_t$ by space-like
hypersurfaces ({\em ie.} such that $\Sigma_t = \{ t=const \}$), and
let $r$ be a radial coordinate which has the property that the
two-surfaces $S_{t,r}$ of constant $t$ and $r$ are approximate metric
spheres with area $4\pi r^2$ for large $r$. We assume that the outer
boundary is described, for each $t \geq 0$, by the two-surface
$S_{t,R}$. Let $n^a$ be the future-directed unit normal to the
surfaces $\Sigma_t$ and let $s^a$ be the normal to the surfaces
$S_{t,r}$ tangent to $\Sigma_t$. Finally, let $v^a$ and $w^a$ be two
mutually orthogonal unit vector fields which are normal to $n^a$ and
$s^a$, and define the real null vector $l^a = (n^a + s^a)/\sqrt{2}$
and the complex null vector $m^a = (v^a + i w^a)/\sqrt{2}$. Then, for
each $L \geq 1$, the boundary condition ${\cal B}_L$ is
\begin{equation}
\frac{\partial}{\partial t} \left.\left[ r^2 l^a\nabla_a \right]^{L-1}
\left( r^5\Psi_0 \right) \right|_{r=R} = 0,
\label{Eq:ImprovedBCNonLinear}
\end{equation}
where $\nabla_a$ denotes the covariant derivative with respect to the
four metric and where, in terms of the Weyl tensor $C_{abcd}$, the
Newman-Penrose scalar $\Psi_0$ is given by $\Psi_0 = C_{abcd} l^a m^b
l^c m^d$. For $L=1$, this reduces to the freezing $\Psi_0$ boundary
condition proposed in
\cite{lKlLmSlBhP05,oSmT05,gNoS06,lLmSlKrOoR06,oR06}. The new boundary
conditions (\ref{Eq:ImprovedBCNonLinear}) are local in time and space,
and do not depend on the spherical harmonic decomposition. Although
they require higher order derivatives of the fields at the boundary,
high-order derivatives can be eliminated by introducing auxiliary
variables at the boundary (for example, see Ref. \cite{dG01}).

Additionally, we have calculated reflection coefficients which
quantify the amount of spurious radiation reflected into the
computational domain both by our new hierarchy of boundary conditions
${\cal B}_L$ together with constraint-preserving boundary conditions
(CBPC), and by CPBC currently in use, which freeze the Newman-Penrose
scalar $\Psi_0$ to its initial value. Including corrections for
backscatter, our new boundary conditions, although no longer perfectly
absorbing, give a reflection coefficient for odd-parity quadrupolar
radiation which is less than the one for the freezing $\Psi_0$
condition by a factor of $M/R$ for $k R > 1.04$. (We expect a similar
result to hold for even-parity quadrupolar radiation.)

An application of our results to simulations of the full nonlinear
Einstein equations requires that: (i) the spacetime near the outer
boundary of the computational domain be accurately described by the
linearized field equations, (ii) the cross sections of the outer
boundary surface with the foliation $\Sigma_t$ be approximate metric
two-spheres of constant area, (iii) the foliation $\Sigma_t$ near the
outer boundary resemble the $t=const.$ foliation of Minkowksi space,
where $t$ is the standard Minkowski time coordinate, and (iv) the
magnitude of the normal component of the shift vector at the outer
boundary be small compared to one. Criteria (i) and (ii) are fully
justified because modern numerical relativity codes can push the outer
boundary into the weak field regime by using mesh refinement, and can
handle spherical outer boundaries using multi-block finite
differencing \cite{jT04,lLoRmT05,eSpDeDmT06} or pseudo-spectral
methods \cite{KST,sBeGpG04}. However, criteria (iii) and (iv) are more
restrictive because they place requirements on the coordinate and
slicing conditions. For example, using maximal slicing or a slicing
which insures that the mean curvature rapidly decays to zero as one
approaches the outer boundary might justify criterion (iii), while
forcing the normal component (with respect to the outer boundary) of
the shift vector to be zero at the outer boundary guarantees (iv). On
the other hand, these criteria are not justified if hyperboloidal
slices are used \cite{jFtV05,sHcStVaZ05,gCcGdH05,lBjB05}, where the
mean curvature asymptotically approaches a constant, nonzero value. It
should not be difficult to generalize our analysis to more general
foliations of Minkowski spacetime using the $2+1$ split discussed in
Sect. \ref{SubSect:2+1}.

The new boundary conditions (\ref{Eq:ImprovedBCNonLinear}) constructed
in this article should be useful for improving the accuracy of binary
black hole calculations on finite domains. For example, if the outer
boundary is spherical with area $4\pi R^2$ and $R > 100M$, then the
reflection coefficient for CPBC with freezing $\Psi_0$ is less than
$0.1\%$ for quadrupolar waves with wavelength $100M$ or smaller. Since
the energy flux scales as the amplitude of the wave squared, this
reflected false radiation causes a relative error in the energy flux
calculation for quadrupolar radiation of the order $10^{-6}$ or
less. If one uses the improved boundary condition ${\cal B}_2$
proposed in this article instead of the freezing $\Psi_0$ condition,
then the reflection coefficient is $100 \times$ smaller; {\em ie.}
less than $0.001\%$. Correspondingly, the contribution of reflected
artefactual radiation to the relative error in the energy flux
calculation for odd-parity quadrupolar radiation is below
$10^{-10}$. Finally, the improved boundary conditions presented here
may be useful for minimizing reflections of ``junk'' radiation present
in the initial data.

We would like to conclude by emphasizing two points. The first point
is that the hierarchy of new boundary conditions ${\cal B}_L$ proposed
in this article (\ref{Eq:ImprovedBCNonLinear}) are not restricted to
the Bianchi equations, but can be applied to any formulation of
Einstein's field equations. It is important that they are used in
conjunction with CPBC and suitable boundary conditions which control
part of the geometry of the outer boundary surface (in particular,
they should insure that the outer boundary remains spherical and that
its area does not change too much in time). While these last two types
of boundary conditions depend explicitly on the formulation, condition
(\ref{Eq:ImprovedBCNonLinear}) does not. After all these boundary
conditions have been specified, one still needs to show that the
resulting IBVP is well posed. This issue is one which we will address
elsewhere. The estimates of the reflection coefficients for spurious
gravitational radiation given in this article are valid for any
representation of the Einstein equations which implements the freezing
$\Psi_0$ boundary condition together with CPBC. In particular, they
are directly applicable to the formulations in
\cite{hFgN99,lKlLmSlBhP05,oSmT05,gNoS06,lLmSlKrOoR06,oR06}.

The second point is that our improved boundary conditions may not be
transparent enough to model accurately all physically interesting
scenarios on an unbounded domain. For example, it is likely that even
with our new boundary conditions, one will find an incorrect tail
decay when measuring the decay of solutions at a fixed location near
the outer boundary. The failure of the simple Sommerfeld condition to
correctly simulate tail decays for a spherically symmetric scalar
field about a Schwarzschild black hole was demonstrated numerically in
\cite{eAeBlBrP04}. In fact, the work in \cite{mDiR04} proves
analytically that the boundary conditions in \cite{eAeBlBrP04} lead to
decay which is faster than any power of $1/t$ (whereas the expected
rate of decay is $1/t^3$ \cite{rP72a}). For future work, we plan to
explore ways to improve our new boundary conditions to reproduce
correctly the tail decay. One possibility is to use the work in
\cite{sL04a,sL04b,sL05} to construct boundary conditions which are
perfectly absorbing when backscatter is considered.

\begin{acknowledgments}
It is a pleasure to thank J. Bardeen, L. Lehner, L. Lindblom,
J. Novak, O. Rinne, J. Stewart, S. Teukolsky, and M. Tiglio for
helpful discussions. L.T.B. was supported by a NASA postdoctoral
program fellowship at the Jet Propulsion Laboratory. O.C.A.S. was
supported in part by NSF grant PHY-0099568, by a grant from the
Sherman Fairchild Foundation to Caltech, and by NSF DMS Award 0411723
to UCSD.
\end{acknowledgments}

\appendix
\section{Stability of the absorbing boundary conditions}
\label{App:Stability}

In this appendix, we consider the initial-boundary value problem
\begin{eqnarray}
&& \left[ \partial_t^2 - \partial_r^2
 + \frac{\ell(\ell+1)}{r^2} \right]\phi(t,r) = 0,
\qquad t > 0,\quad R_0 < r < R,
\label{Eq:IBVPEQ}\\
&& \phi(0,r) = f(r), \qquad 
   \partial_t\phi(0,r) = g(r),
\qquad R_0 < r < R,
\label{Eq:IBVPID}\\
&& (b_+)^{L+1}\partial_t\phi(t,R_0) = 0, \qquad
   (b_-)^{L+1}\partial_t\phi(t,R) = 0,
\qquad t > 0,
\label{Eq:IBVPBC}
\end{eqnarray}
where $\ell$, $L$ are natural numbers, $0 < R_0 < R$ are the inner and
outer radii of a spherical shell, $f$ and $g$ are smooth initial data,
and $b_\pm = r^2(\partial_t \mp \partial_r)$. The reason for
introducing the inner boundary at $r = R_0$ is to excise the
coordinate singularity at $r=0$. This is not a restriction for the
purpose of this article, since we are interested only in the region near
the outer boundary, and since we consider the linearized equations
for modeling a physical scenario away from the strong field region.

In order to show that the problem
(\ref{Eq:IBVPEQ},\ref{Eq:IBVPID},\ref{Eq:IBVPBC}) is stable in the
sense that the solution depends continuously on the data, we introduce
the following notation:
\begin{displaymath}
\Phi_j^{(\pm)} = (b_\pm)^j\phi,
\qquad j = 0,1,2,3,...
\end{displaymath}
Notice that $\Phi_0^{(+)} = \Phi_0^{(-)} = \phi$. By applying the
operators $b_+$ and $b_-$ to both sides of Eq. (\ref{Eq:IBVPEQ}), one
finds the formula
\begin{displaymath}
b_{\mp}(\Phi_j^{(\pm)}) =
  \pm 2j r\Phi_j^{(\pm)} - (\ell-j+1)(\ell+j)r^2\Phi_{j-1}^{(\pm)}
\end{displaymath}
for $j=1,2,3,...$. Using this and $b_\pm(\Phi_j^{(\pm)}) =
\Phi_{j+1}^{(\pm)}$, we find that
\begin{displaymath}
2r^2\partial_t( \Phi_j^{(\pm)}) = (b_+ + b_-)\Phi_j^{(\pm)}
 = \pm 2j r\Phi_j^{(\pm)} - (\ell-j+1)(\ell+j)r^2\Phi_{j-1}^{(\pm)}
 + \Phi_{j+1}^{(\pm)}
\end{displaymath}
for all $j=1,2,3,...$. Therefore, the evolution equations
(\ref{Eq:IBVPEQ}) and the boundary conditions (\ref{Eq:IBVPBC}) yield
the evolution system
\begin{eqnarray}
\partial_t\Phi_0 &=& \frac{1}{2r^2}\left( \Phi_1^{(+)} + \Phi_1^{(-)} \right),
\label{Eq:Large1}\\
\partial_t\Phi_j^{(+)} &=& \frac{1}{2r^2} \Phi_{j+1}^{(+)}
 + \frac{j}{r} \Phi_j^{(+)} - \frac{(\ell-j+1)(\ell+j)}{2}\Phi_{j-1}^{(+)},
\qquad j = 1,2,...L,
\label{Eq:Large2}\\
\partial_t\Phi_j^{(-)} &=& \frac{1}{2r^2} \Phi_{j+1}^{(-)}
 - \frac{j}{r} \Phi_j^{(-)} - \frac{(\ell-j+1)(\ell+j)}{2}\Phi_{j-1}^{(-)},
\qquad j = 1,2,...L,
\label{Eq:Large3}\\
(\partial_t + \partial_r)\Phi_{L+1}^{(+)} 
 &=& \frac{2(L+1)}{r} \Phi_{L+1}^{(+)} - (\ell-L)(\ell+L+1)\Phi_{L}^{(+)},
\label{Eq:Large4}\\
(\partial_t - \partial_r)\Phi_{L+1}^{(-)} 
 &=& -\frac{2(L+1)}{r} \Phi_{L+1}^{(-)} - (\ell-L)(\ell+L+1)\Phi_{L}^{(-)},
\label{Eq:Large5}
\end{eqnarray}
with boundary conditions
\begin{equation}
\partial_t\Phi_{L+1}^{(+)}(t,R_0) = 0, \qquad
\partial_t\Phi_{L+1}^{(-)}(t,R) = 0, \qquad t > 0.
\label{Eq:LargeBC}
\end{equation}
The system
(\ref{Eq:Large1},\ref{Eq:Large2},\ref{Eq:Large3},\ref{Eq:Large4},\ref{Eq:Large5})
constitutes a symmetric hyperbolic system with maximally dissipative
boundary conditions (\ref{Eq:LargeBC}). It is well-known (see, for
example, Ref. \cite{KL89}) that such systems are well posed and admit
energy estimates. For example, it follows that a smooth enough
solution satisfies the estimate
\begin{equation}
E(t) \leq e^{b t} E(0)
\label{Eq:EnergyEstimate}
\end{equation}
with the energy norm
\begin{displaymath}
E(t) = \frac{1}{2}\int\limits_{R_0}^R \left( r^{2(L+1)}\Phi_0^2(t,r) 
+ \sum\limits_{j=1}^{L+1} r^{2(L+1-j)}
  \left[ (\Phi_j^{(+)}(t,r))^2 + (\Phi_j^{(-)}(t,r))^2 \right]
\right) dr,
\end{displaymath}
where $b$ is a constant that does not depend on the solution. In
particular, the inequality (\ref{Eq:EnergyEstimate}) implies that the
solutions depend uniquely and continuously on the initial data. The
existence of solutions (including for evolution equations with more
general potentials than the one in Eq. (\ref{Eq:IBVPEQ})) can be
proved using methods from semigroup theory; see for example chapter
6.3 in \cite{hB05} for a well posedness proof for a similar problem.
A different well posedness proof based on the verification of the
Kreiss condition is given in \cite{aBeT80}.

Since for $\ell \leq L$, the exact outgoing solutions
$\phi_{\nearrow,\ell}(t,r)$ constructed in
Sect. \ref{SubSect:ExactSolutions} satisfy the boundary conditions
(\ref{Eq:IBVPBC}), provided the function $U_\ell$ is compactly
supported in $(R_0,R)$, it follows that the boundary conditions
(\ref{Eq:IBVPBC}) are perfectly absorbing for all $\ell \leq L$.

\bibliography{refs}
\end{document}